\documentclass[12pt,journal]{IEEEtran}
\usepackage{epsfig,makeidx,color}
\usepackage{graphicx}
\usepackage{amsbsy}
\usepackage{amsmath}
\usepackage{amssymb}
\usepackage{euscript}

\newtheorem{mytheorem}{\bf Theorem}[section]

\newcommand{\ave}{{\mathbb E}}

\newcommand {\Define} {\stackrel {\Delta} {=}  }

\renewcommand{\baselinestretch}{1.775}
\begin{document}

\title{Modulation Diversity in Fading Channels with Quantized Receiver}
\author{Saif~Khan~Mohammed*,~\IEEEmembership{Member,~IEEE,}\\
        Emanuele~Viterbo,~\IEEEmembership{Fellow,~IEEE}
        Yi~Hong,~\IEEEmembership{Senior~Member,~IEEE,} \\
        and~Ananthanarayanan Chockalingam,~\IEEEmembership{Senior~Member,~IEEE}
\thanks{\scriptsize Saif K. Mohammed is with the Communication Systems Division at the Dept. of Electrical
Eng. (ISY) Link{\"o}ping University, 581 83 Link{\"o}ping, Sweden. E-mail: $\tt saif$@$\tt isy.liu.se$.}
\thanks{\scriptsize Emanuele Viterbo and Yi Hong are with the Dept. of Electrical and Computer Systems Eng.
Monash University at Clayton, Melbourne, Victoria 3800, Australia. E-mail: $\tt \{emanuele.viterbo, yi.hong\}$@$\tt monash.edu$.}
\thanks{\scriptsize A. Chockalingam is with the Dept. of Electrical and Communication Eng. (ECE) Indian Institute of Science,
Bangalore $560012$, India. E-mail: $\tt achockal$@$\tt ece.iisc.ernet.in$.}
}

\onecolumn
\maketitle


\begin{abstract}
In this paper, we address the design of codes which achieve modulation diversity
in block fading single-input single-output (SISO) channels
with signal quantization at receiver and low-complexity decoding.
With an unquantized receiver, coding based on algebraic rotations is known
to achieve modulation coding diversity. On the other hand, with a quantized receiver,
algebraic rotations may not guarantee diversity.
Through analysis, we propose specific rotations which result in
the codewords having equidistant component-wise projections.
We show that
the proposed coding scheme achieves
maximum modulation diversity with a low-complexity minimum distance decoder and perfect channel knowledge.
Relaxing the perfect channel knowledge assumption we propose a novel training/estimation and receiver control
technique to estimate the channel. We show that our coding/training/estimation scheme and minimum distance decoding achieve an error probability performance similar to that achieved with perfect channel knowledge. 
\end{abstract}
\IEEEpeerreviewmaketitle

\section{Introduction}
In practical communication receivers, the analog received signal is quantized into a finite number of bits for
further digital baseband processing. With increasing bandwidth requirements of
modern communication systems, analog-to-digital converters (ADC) are required to operate at high frequencies.
However, at high operating frequencies, the precision of ADC's is limited \cite{Walden}. Limited precision generally leads to high quantization noise, which degrades performance.  In case of fading channels, floors in the bit
error performance have been reported, and it seems difficult to avoid this behavior \cite{Nossek}\cite{Gareth}. On the other hand, channel capacity results show that even with 2-bit quantizers, the capacity of a quantized output channel is not far from that of a channel with unquantized output \cite{Ivrlac}\cite{Singh}. Therefore, there appears to be a gap between the theoretical limits
of communication with quantized receivers, and the current state of art.

In communication systems with fading, an important performance metric
is the reliability of reception.
For single antenna fading scenarios, modulation diversity is a well known signal space diversity technique
to improve the reliability/diversity of reception \cite{Vit98}\cite{Vit96}. However, with a quantized receiver, this coding alone {\em does not} guarantee
improvement in diversity.

In this paper, we propose 2-dimensional constellations rotated by an angle $\theta$
which can achieve full modulation diversity with a quantized receiver.
With a quantized receiver, the maximum likelihood (ML) decoder is not the usual minimum distance decoder, and would be much more complex to implement. We therefore assume a minimum distance decoder operating on the quantized
channel outputs. We observe that, with a quantized receiver, {\em i}) for a given rate of information transmission in bits per channel use,
there is a minimum requirement on the number of quantization bits, without which floors\footnote{\footnotesize {Error probability performance is said to {\em floor},
if and only if it converges to a non-zero positive constant as the signal-to-noise ratio tends to infinity.}} appear in the error probability performance, {\em ii}) there is only a small subset of {\em admissible} rotation angles
which can guarantee diversity improvement and no error floors, and {\em iii}) for a quantized receiver with perfect channel knowledge and minimum distance decoding, we analytically show that, among all {\em admissible} rotation angles,
a good choice is one in which the transmitted vectors have {\em equidistant projections} along both the transmitted components.
We then show that the square $M^2$-QAM constellation rotated by $\theta = \tan^{-1}(1/M)$ has equidistant projections.

Further, we relax the perfect channel knowledge assumption, and propose
novel training sequences and channel estimation scheme, which
achieve an error probability performance close to that achieved
with perfect channel knowledge.
Through Monte-Carlo simulations we show that even with coarse
analog-to-digital conversion, and short training sequences, the error performance with the estimated channel is similar to that with perfect channel knowledge.
The main {\em interesting} result is that, even when the channel estimate is not perfect,
an error probability performance exactly same as that with perfect channel estimate is achievable
under some sufficiency conditions on the channel estimate and the number of quantization bits. These conditions are analytically derived,
and shown
to be satisfied by the proposed training/estimation scheme for some scenarios.
Another interesting result is that, with sufficient number of quantization bits, the error performance never floors {\em irrespective} of the
quality of the channel estimate.   
\section{System model and Quantized Receiver} \label{sec2}
We consider SISO block fading channels with single transmit and single receive antenna.
The channel gains are assumed to be quasi-static for the coherence interval of the channel, and change to an independent realization in the next coherence interval. We further assume that the signaling bandwidth is much smaller than the
coherence bandwidth of the channel (frequency flat fading), and therefore the channel frequency
response is assumed to have constant magnitude and linear phase within the signalling
bandwidth. Let the radio frequency band used for transmission be $(f_c - W/2 , f_c + W/2)$, where $f_c \gg W$
is the carrier frequency and $W$ is the signaling bandwidth. The complex channel frequency response is then given by
\begin{equation}
\label{channel_resp}
H(f) = \vert h \vert e^{-j2\pi \tau f}\,,\, \vert f - f_c \vert \leq \frac{W}{2},
\end{equation}
and zero elsewhere, i.e., scaling by $\vert h \vert$, and a delay of $\tau$ seconds.
The transmitted signal is given by
\begin{equation}
\label{xt_eq}
x(t) = \sum_{k} ( x_k^I \cos(2\pi f_c t) + x_k^Q \sin(2 \pi f_c t)) g(t - kT),
\end{equation}
where $1/T$ is the rate at which information symbols are transmitted, and $x_k = x_k^I + j x_k^Q$ is the $k$-th transmitted
information symbol.
We assume pulse shaping signals which result in no inter-symbol interference (ISI) (e.g., $g(t) = \mbox{sinc}(Wt)$).
Prior to the transmission of $K$ information symbols, there is a training phase
in which a known preamble sequence of $P$ symbols is transmitted to enable carrier frequency synchronization in the receiver (i.e., enabling the phased locked loop (PLL) in the receiver to lock to the
transmitter's local oscillator) and also for tuning the receiver gain.
In this paper we assume the preamble to be a constant amplitude carrier obtained by setting $x_k^I = A$ and $x_k^Q = 0$ in (\ref{xt_eq}).
The received signal during the training phase of duration $P$ symbols, is given by $y(t) = A \vert h \vert \cos(2\pi f_c t - 2 \pi f_c \tau) \sum_{k=0}^{P-1} g( (t - \tau) - kT)$.
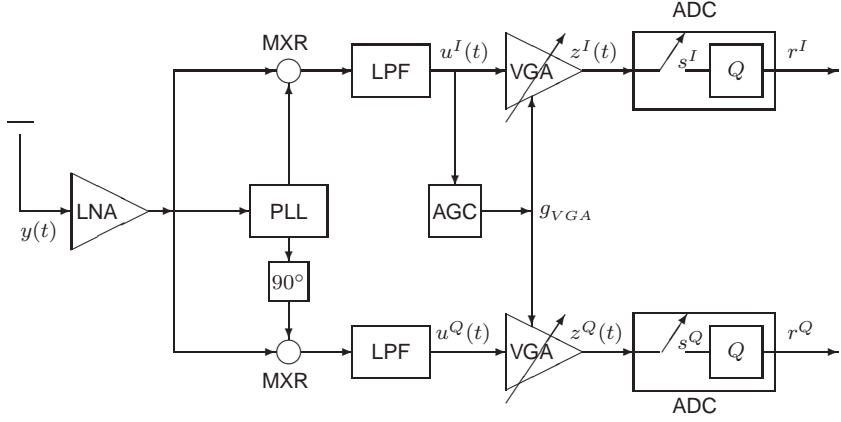
\begin{figure}[t]
\begin{center}
\setlength{\unitlength}{3.4mm}
\begin{picture}(34,16)(1,0)
\scriptsize

\put(11.5,2.5){\circle{0.9}}
\put(11.5,13.5){\line(1,1){0.2}}  \put(11.5,13.5){\line(1,-1){0.2}}
\put(11.5,13.5){\line(-1,1){0.2}} \put(11.5,13.5){\line(-1,-1){0.2}}
\put(11.5,13.5){\circle{0.9}}
\put(11.5,2.5){\line(1,1){0.2}}  \put(11.5,2.5){\line(1,-1){0.2}}
\put(11.5,2.5){\line(-1,1){0.2}} \put(11.5,2.5){\line(-1,-1){0.2}}

\put(1,8){\vector(1,0){2}}   \put(1,8){\line(0,1){3}}
\put(1,11){\line(1,1){0.5}}  \put(1,11){\line(-1,1){0.5}}  \put(0.5,11.5){\line(1,0){1}}

\put(3,7){\makebox(2,2){\sf LNA}}
\put(3,6.5){\line(0,1){3}} \put(3,6.5){\line(2,1){3}} \put(3,9.5){\line(2,-1){3}}

\put(10,7){\framebox(3,2){\sf PLL}}
\put(10.75,4.5){\framebox(1.5,1.5){$90^\circ$}}
\put(6,8){\vector(1,0){1}}  \put(7,8){\vector(1,0){3}}
\put(7,2.5){\line(0,1){11}} \put(7,2.5){\vector(1,0){4}}
\put(7,13.5){\vector(1,0){4}}
\put(11.5,7){\vector(0,-1){1}}
\put(11.5,4.5){\vector(0,-1){1.5}}
\put(11.5,9){\vector(0,1){4}}

\put(12,13.5){\vector(1,0){2}}      \put(12,2.5){\vector(1,0){2}}
\put(14,12.5){\framebox(3,2){\sf LPF}}  \put(14,1.5){\framebox(3,2){\sf LPF}}
\put(17,13.5){\vector(1,0){3}}      \put(17,2.5){\vector(1,0){3}}
\put(17,7){\framebox(2,2){\sf AGC}}
\put(19,8){\vector(1,0){2}}         \put(18,13.5){\vector(0,-1){4.5}}

\put(20,12.5){\makebox(2,2){\sf VGA}}
\put(20,12){\line(0,1){3}} \put(20,12){\line(2,1){3}} \put(20,15){\line(2,-1){3}}
\put(20,1.5){\makebox(2,2){\sf VGA}}
\put(20,1){\line(0,1){3}} \put(20,1){\line(2,1){3}} \put(20,4){\line(2,-1){3}}

\put(21,8){\vector(0,-1){4.5}}      \put(21,8){\vector(0,1){4.5}}
\put(20,11.5){\vector(2,3){2.3}}    \put(20,0.5){\vector(2,3){2.3}}
\put(23,13.5){\vector(1,0){2}}      \put(23,2.5){\vector(1,0){2}}

\put(26,13.5){\vector(2,3){1}}    \put(26.1,2.5){\vector(2,3){1}}

\put(25,13.5){\line(1,0){1}}      \put(25,2.5){\line(1,0){1}}
\put(27,13.5){\line(1,0){1}}      \put(27,2.5){\line(1,0){1}}
\put(27,13.5){\line(-2,1){1}}      \put(27,2.5){\line(-2,1){1}}

\put(25,12){\framebox(5.5,3){}}     \put(25,1){\framebox(5.5,3){}}
\put(28,12.5){\framebox(2,2){$Q$}}  \put(28,1.5){\framebox(2,2){$Q$}}
\put(30,13.5){\vector(1,0){3}}      \put(30,2.5){\vector(1,0){3}}

\put(1.0,7){$y(t)$}
\put(17.4,14){$u^I(t)$}  \put(17.25,3){$u^Q(t)$}
\put(22.5,14){$z^I(t)$}  \put(22.5,3){$z^Q(t)$}
\put(31,14){$r^I$}  \put(31,3){$r^Q$}
\put(26.7,13.6){$s^I$}  \put(26.7,2.6){$s^Q$}

\put(21.3,7.9){$g_{_{VGA}}$}
\put(10.45,14.3){{\sf MXR}}
\put(10.45,1.1){{\sf MXR}}
\put(26.5,15.6){{\sf ADC}}
\put(26.5,0.1){{\sf ADC}}

\end{picture}
\end{center}
\vspace{-3mm}
\caption{Receiver analog front end (AFE) .}
\vspace{-5mm}
\label{qmimo_fig}
\end{figure}
Figure \ref{qmimo_fig} shows the signal path of the analog front end of
a typical heterodyne receiver \cite{QizhengGu}.
Let the combined gain of the Low Noise Amplifier (LNA), Mixer (MXR) and Low Pass Filter (LPF) be denoted by $g_{AFE}$.
In the training phase, after the PLL has locked, the LPF output (I Path) is given by $u^{I}(t) = A g_{AFE} \vert h \vert  \sum_{k=0}^{k = P-1} g((t - \tau) - kT) + n^{I}(t)$,
where $n^{I}(t)$ is the white Gaussian noise in the receiver (I path).
The LPF output is digitized using a Nyquist rate sample \& hold type analog-to-digital converter (ADC), as shown in Fig. \ref{qmimo_fig}.
Let the input dynamic range of the ADC be $-c_q/2$ to $c_q/2$. We also refer to $c_q/2$ as the clip level, since any input greater than $c_q/2$ would be limited to $c_q/2$. For optimum performance, it is desirable that the range of the input signal to the ADC matches with the ADC dynamic range (ADC range matching). Due to fading, the input level at the ADC
may vary, and therefore a variable gain amplifier (VGA) is generally used to ensure ADC range matching. The gain of the VGA is controlled by the automatic gain control (AGC) module \cite{QizhengGu}. During the training phase, the
AGC detects the peak of the signal $u^{I}(t)$ using a conventional analog peak detector whose output is given by
\begin{equation}
\label{preamble_agc_lpf_out}
V_{agc-pk} = A g_{AFE} \vert h \vert.
\end{equation}
Let $X$ denote the peak absolute value of the transmitted symbols, $x_k^I$ and $x_k^Q$, during normal information transmission phase. 
During information transmission phase, ADC range matching (i.e., $\frac{c_q}{2} ~=~ g_{VGA} g_{AFE} \vert h \vert X$) requires the VGA gain to be
\begin{eqnarray}
\label{g_vga_eqn}
g_{VGA} = \frac{c_q}{2} \frac {A} {X} \frac {1}{V_{agc-pk}}.
\end{eqnarray}
Since the ratio $A/X$ and $c_q/2$ are known {\em a priori}, this computation is done in the AGC using simple analog circuits \cite{Liu}.
In the rest of the paper, we assume that this computation is perfect.

During the information transmission phase, the PLL tracking loop is turned off and
the VGA gain setting is frozen to the value given by (\ref{g_vga_eqn}).
Therefore, during this phase, the ADC input signal (I path) is given by
$z^{I}(t) = g_{VGA} (g_{AFE} \vert h \vert \sum_{k=P}^{K+P-1} x_k^I g(t - \tau - kT) + n^{I}(t))
= \frac {c_q}{2} \sum_{k=P}^{K+P-1} \frac {x_k^I} {X} g(t - \tau - kT) + g_{VGA} n^{I}(t)$.
The ADC input (Q-path) is similar.
Subsequently, without loss of generality, we assume an ADC with a normalized clip level of $c_q/2 = 1$. Assuming perfect timing synchronization (i.e., receiver can perfectly estimate $\tau$), the $k$-th output of the sample \& hold circuit, at time $t = \tau + kT$ is given by
\begin{eqnarray}
\label{quantizer_input}
s^{I}_k = \frac {x_k^I} {X} +  \frac {w^{I}_k} {\vert h \vert X}~~,~~
s^{Q}_k = \frac {x_k^Q} {X} +  \frac {w^{Q}_k} {\vert h \vert X},
\end{eqnarray}
where $w^{I}_k\Define{n^{I}(\tau + kT)}/{g_{AFE}}$ and $w^{Q}_k\Define{n^{Q}(\tau + kT)}/{g_{AFE}}$
are i.i.d. Gaussian random variables with variance denoted by $\sigma^2/2$.
Let the average
transmit power be denoted by $P_T \Define \ave[\vert x_k \vert^2 ]$.
Then the instantaneous signal to noise ratio (SNR) at the output of the sample \& hold circuit is given by $\gamma_{inst} \Define {P_T \vert h \vert^2}/{\sigma^2}$.
Assuming a Rayleigh fading model with $h \sim {\mathcal C}{\mathcal N}(0,1)$ (Complex Gaussian with mean zero and variance $1$), the average signal to noise ratio (SNR) is given by $\gamma \Define \ave_{h} [ \gamma_{inst} ] = {P_T}/{\sigma^2}$.
The output of the sample \& hold circuit is then quantized by a $b$-bit uniform quantizer $Q$, as shown in Fig. \ref{qmimo_fig}.
The quantizer is modeled by the function $Q_{b}(t), t \in {\mathbb R}$, which is given by
\begin{eqnarray}
\label{qri}
Q_b(t) &=& \left\lbrace
\begin{array}{cc}
+1, & \hspace{-3mm} \xi(t)  \geq  (2^{b - 1} - 1) \\
-1, & \hspace{-1mm} \xi(t)  \leq  -(2^{b - 1} - 1) \\
\frac {(2\xi(t) + 1)} { 2^{b} - 1 },  & {\text{otherwise}} \\
\end{array}\right. \\
& &\xi(t) \Define {\Big \lfloor} \frac { t \, (2^{b} - 1)}{2} {\Big \rfloor }
\end{eqnarray}
where $\lfloor x \rfloor$ denotes the largest integer not greater than $x$.
For a $n$-dimensional complex vector ${\bf z} = (z_1, z_2, \cdots, z_n)$, let ${\bf Q}_b({\bf z})$ denote the $n$-dimensional component-wise quantized version of ${\bf z}$. That is, ${\Tilde {\bf z}} = (  {\Tilde z}_1, {\Tilde z}_2, \cdots, {\Tilde z}_n) = {\bf Q}_b({\bf z})$ implies that
\begin{eqnarray}
\label{zq}
{\Tilde z}_{i}^{I} = Q_b(z_i^{I})\,\,,\,\,{\Tilde z}_{i}^{Q} = Q_b(z_i^{Q})\,\,\, i=1,2, \ldots ,n.
\end{eqnarray}
The $k$-th quantized received symbol, $r_k = r_k^I + j r_k^Q$ is therefore given by
\begin{equation}
\label{eqn_rk}
r_k^I = Q_b(s_k^I)\,,\,r_k^Q = Q_b(s_k^Q)
\end{equation}
where $s_k^I$ and $s_k^Q$ are the real and imaginary components of the $k$-th sample \& hold output symbol. 

\begin{figure}[t]
\begin{center}
\hspace{-7mm}
\epsfig{file=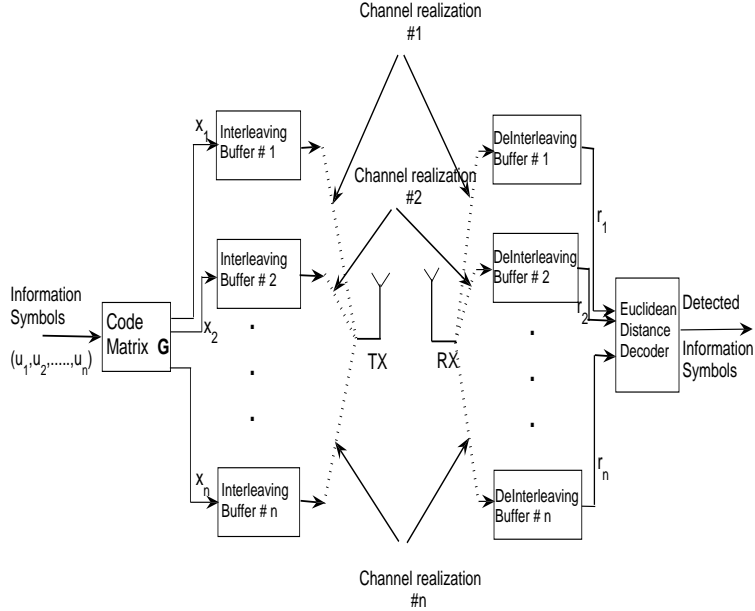, width=100mm,height=85mm}
\end{center}
\vspace{-5mm}
\caption{Achieving modulation diversity by coding across $n$ different channel realizations.}
\vspace{-5mm}
\label{tx_rx_fig}
\end{figure}

Modulation diversity coding is illustrated in Fig.~\ref{tx_rx_fig}.
Coding is performed across $n > 1$ information symbols resulting in $n$ coded symbols/codeword.
These $n$ coded symbols are interleaved and then transmitted over $n$ independent channel coherence intervals (realizations).
At the receiver, the channel outputs during the $n$ coherence intervals are buffered, followed by de-interleaving and detection.
Suitable coding across $n$ independent channel realizations results in an $n$-fold increase in the diversity of reception. In fading channels, codes designed using algebraic lattices can achieve modulation diversity, and are therefore employed to improve the diversity of reception \cite{Vit98}.
With an unquantized receiver, it is known that
lattice codes based on algebraic rotations can achieve full modulation diversity \cite{Vit96}\cite{Vit04}. {\em However, with
quantized receivers, this is no longer true}.
In this paper we consider the case of $n=2$. Let the information symbol vector be denoted by ${\bf u} = (u_1, u_2)^T$, where the information symbols $u_1$ and $u_2$ are restricted to square $M^2$-QAM signal set, though a generalization to non-square QAM is trivial. Let the set ${\mathcal S}_M = \{ -(M -1), \ldots ,-1,1, \cdots ,(M-1) \}$ denote the $M$-PAM
signal set. Then, $M^2$-QAM is denoted by the set ${\mathcal S}_M^2 \Define \{ w + j v ~|~ w,v \in {\mathcal S}_M \}$. The information symbols are coded using a $2 \times 2$ rotation matrix ${\bf G}$, resulting in the transmit vector ${\bf x} = (x_1 , x_2)^T={\bf G} {\bf u}$, where
\begin{eqnarray}
\label{modcod}
{\bf G} = \left[\begin{array}{cc}
\cos(\theta) & \sin(\theta) \\
-\sin(\theta) &  \cos(\theta)
\end{array} \right].
\end{eqnarray}
Due to QAM symmetry, one can restrict the rotation angle in (\ref{modcod}) to $[0, \pi/4)$.   
The set of transmitted vectors ${\mathcal X}$ and the peak component value $X$ are given by
\begin{eqnarray}
\label{xmt_set}
{\mathcal X} = {\Bigg \{}  {\bf x} ~~|~~ {\bf x} = {\bf G} {\bf u}, {u_1},{u_2} \in {\mathcal S}_M^2 {\Bigg \}} \,\,,\,\,
X = \max_{{\bf x} \in  {\mathcal X}} {\Bigg \{} \max_{i = 1,2} {\Big [}  \max(\vert x_i^I \vert, \vert x_i^Q \vert) {\Big ]} {\Bigg \}}
\end{eqnarray}
Also, let the channel gain during the transmission of $x_1$ and $x_2$ be denoted by $\vert h_1 \vert$ and
$\vert h_2 \vert$, respectively.
We assume $h_1$ and $h_2$ to be i.i.d. ${\mathcal C}{\mathcal N}(0,1)$.
Let ${\bf r} = (r_1 , r_2)^T$ denote the quantized received vector, where $r_1=r_1^{I} + jr_1^{Q}$ and $r_2=r_2^{I} + jr_2^{Q}$ are the ADC outputs during the transmission of $x_1$ and $x_2$, respectively. From (\ref{quantizer_input}) and (\ref{eqn_rk}) it follows that
\begin{equation}
\label{rcv_vec}
r_i^{I} = Q_b{\Big (}  \frac {x_i^I} {X} +  \frac {w^{I}_i} {\vert h_i \vert X} {\Big )}\,,\,
r_i^{Q} = Q_b{\Big (}  \frac {x_i^Q} {X} +  \frac {w^{Q}_i} {\vert h_i \vert X} {\Big )}.
\vspace{-3mm}
\end{equation}
With the above quantized receiver model, maximum likelihood decoding
is no more given by the minimum distance decoder, and is rather
complex. Nevertheless, due to its lower decoding complexity,
we shall assume a minimum
distance decoder taking ${\bf r}$ as its input,
and the output (detected information symbols) given by
\begin{eqnarray}
\label{euc_dis_dec}
{\widehat {\bf u}} & = & \arg \hspace{-3mm} \min_{{\bf u} \in {{\mathcal S}_M^2} \times {{\mathcal S}_M^2}} {\Big \Vert} \mbox{diag}(\vert h_1 \vert,\vert h_2 \vert) {\Big (} {\bf r} - \frac{{\bf G}{\bf u}}{X} {\Big )} {\Big \Vert}^2 \nonumber \\
& = & \arg \hspace{-3mm} \min_{{\bf u} \in {{\mathcal S}_M^2} \times {{\mathcal S}_M^2}} {\Bigg (} {\bf r} - {\Big (} \frac{{\bf G}{\bf u}}{X} {\Big )} {\Bigg )}^{\dag} {\bf D}_{\rho} {\Bigg (} {\bf r} - {\Big (} \frac{{\bf G}{\bf u}}{X} {\Big )} {\Bigg )}  \,\,,\,\, {\bf D}_{\rho} \Define \mbox{diag}( 1 , \rho^2)
\end{eqnarray}
where $\rho \Define \vert h_2 \vert/\vert h_1 \vert$ is the channel gain ratio, and
$\dag$, $\Vert . \Vert$ denote Hermitian transpose and Euclidean norm respectively.
In the subsequent sections \ref{sec3} and \ref{code_design}, assuming perfect receiver knowledge of $\rho$, we show that even with the suboptimal minimum distance decoder in (\ref{euc_dis_dec}),
we can avoid error floors and also achieve modulation diversity.
Finally in section \ref{imperfect_rho}, we relax the perfect channel knowledge
assumption and present a practical training scheme to
estimate $\rho$. We show that even with coarse quantization (i.e., small number of quantization bits -- $b$),
the proposed channel estimation scheme achieves error probability performance close to that achieved with perfect knowledge of $\rho$.
\section{Rotation Coding in Quantized Receiver}
\label{sec3}
In an unquantized receiver, at high SNR,
the word error probability is minimized by choosing the transmit vectors such that
the minimum product distance between any two vectors is maximized \cite{Vit96}.
There also exists algebraic rotations which guarantee a non-vanishing minimum product
distance with increasing QAM size \cite{Vit98}.
In this paper, we study the error performance of these rotated constellations with a {\em quantized} receiver and minimum distance decoding, and derive the conditions under which full modulation diversity can be achieved.

\begin{figure}[t]
\begin{center}
\hspace{-7mm}
\epsfig{file=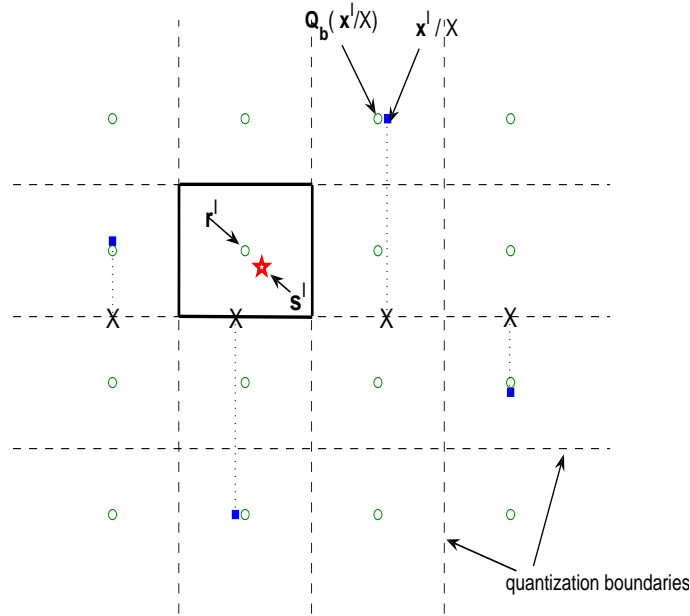, width=90mm,height=90mm}
\end{center}
\vspace{-5mm}
\caption{Signal space at the quantizer input with $b=2$ (real component). Rotated $4$-QAM ($\theta=20^{\circ}$) depicted with dark filled squares.}
\vspace{-5mm}
\label{siso_code_fig}
\end{figure}

In case of a quantized receiver, the sample \& hold outputs (\ref{quantizer_input}), are quantized to the appropriate {\em quantization box} containing it.
As an example, Fig. \ref{siso_code_fig} illustrates the rotated $4$-QAM constellation with $\theta = 20^{\circ}$. The dark filled squares represent the 4 possible values taken by the real component of the normalized transmit vector ${\bf x}^{I}/X = ( x_1^{I}/X \,,\, x_2^{I}/X)^T$.
The {\em projections} of the
$4$ possible vectors onto the first component (horizontal) are marked with a cross.
A $b=2$-bit quantizer is used along both codeword components. The dashed horizontal and vertical lines
represent the quantization boundaries along the 2 components.
As an example, in Fig.\ref{siso_code_fig} the real component of the sample \& hold output vector ${\bf s}^{I}=(s_1^{I}, s_2^{I})^T$ (marked with a star),
is therefore quantized to ${\bf r}^{I}=(r_1^{I}, r_2^{I})^T$ (Note that there are totally 16 different quantized outputs marked with empty circles).
The quantization box corresponding to the output ${\bf r}^{I}$ is shown in the figure as a square
with solid lines.

As the noise variance $\sigma^2
\rightarrow 0$, the sample \& hold output ${\bf s}$ is almost the same as the normalized transmitted vector ${\bf x}/{X}$.
Therefore at sufficiently high SNR, if there exists two different transmit vectors ${\bf x}$ and ${\bf y}$,
such that ${\bf Q}_b({\bf x}/{X})$ and
${\bf Q}_b({\bf y}/{X})$ are identical, then it is obvious that the error probability performance would floor
as SNR $\rightarrow \infty$. This is because, at high SNR the quantizer output would be the same irrespective of whether ${\bf x}$ or ${\bf y}$ was 
transmitted, which makes it impossible for the
the receiver to distinguish between the two transmit vectors leading to erroneous detection. More formally, two transmit vectors ${\bf x}$ and ${\bf y}$
are said to be {\em distinguishable} if and only if ${\bf Q}_b({\bf x}/{X}) \ne {\bf Q}_b({\bf y}/{X})$.
Therefore, in order to avoid floors in the error probability performance, we propose the first code design criterion.

{\em {\bf Criterion I} : A necessary and sufficient condition to avoid error floors with
a quantized receiver, is that any two transmit vectors must be distinguishable.}

To achieve full modulation diversity, it is required that even under deep fading conditions in
one component, any two transmit vectors ${\bf x}$ and ${\bf y}$
must still be {\em distinguishable} in the other component.
This, therefore, implies that the
projections of all the transmit vectors onto
any one component must be {\em distinguishable} by the quantizer in that component.
Therefore, we have the second criterion.

{\em {\bf Criterion II} : Given a $b$-bit quantized receiver, in order to achieve full modulation diversity, a necessary condition on the rotation angle $\theta$ is that, any two distinct transmit vectors ${\bf x}$ and ${\bf y}$ satisfy
\begin{eqnarray}
\label{crit_2}
{\bf Q}_b(x_i / X) \ne {\bf Q}_b(y_i / X), \,\,\, i=1,2.
\end{eqnarray}
}
With a rotated $M^2$-QAM there are totally
$M^2$ distinct projections onto any component, and therefore the minimum number
of quantization bits required for the transmit vectors to be {\em distinguishable} along any component is at least $\lceil {2 \log_2(M)} \rceil$. Hence, in order to achieve full modulation diversity 
a straight forward lower bound on $b$ is\footnote {\footnotesize{$\lceil x \rceil$ denotes the smallest integer not smaller than $x$.}}
\begin{equation}
\label{Qb_lb_sisoM}
b \geq \lceil {2 \log_2(M)} \rceil.
\end{equation}
Subsequently, we assume that for a given $M$, $b$ is fixed to the lower bound value in (\ref{Qb_lb_sisoM}).
We further note that, with a $b = \lceil {2 \log_2(M)} \rceil$-bit quantizer, {Criterion II} is not satisfied
by all rotation angles\footnote{\footnotesize {For example, even though $\theta = 1/2\tan^{-1}(2)$ guarantees a rotation code having non-vanishing minimum
product distance, with a $b=4$-bit uniform quantizer and $M^2=16$-QAM it {\em does not} satisfy {Criterion II}.}}.

With a $b = \lceil 2\log_2(M) \rceil$-bit quantizer, the set of angles (between 0 and $\pi/4$) which result in {\em distinguishable} projections along both the codeword components will be referred to as the {\em admissible} angles (i.e., angles which satisfy {Criterion II}).
For example, with 4- and 16-QAM, the admissible angles lie in the range $(\tan^{-1}(1/5) \,\,\, \pi/4)$ and
$(11.3^{\circ} \,\,\, 16.9^{\circ})$, respectively.
With increasing $M$, the interval of {admissible} angles reduces. With 256-QAM, the range of {admissible} angles is only $(3.47^{\circ} \,\,\, 3.68^{\circ})$. Another interesting fact is
that, for $M^2$-QAM, $\theta = \tan^{-1}(1/M)$ is always in the set of {admissible} angles.
Further, as $M$ increases, $\tan^{-1}(1/M) \pm \epsilon$ are observed to be the only {admissible} angles.

Apart from the fact that the chosen angle must have distinguishable projections,
it can be analytically shown that for $M^2$-QAM, any rotation angle for which
the rotated constellation satisfies
\begin{equation}
\label{matched_constel}
{\bf Q}_b({\bf x}/X) = {\bf x}/X\,\,,\,\, {\bf x} \in {\mathcal X}
\end{equation}
does indeed achieve a diversity order of 2 (i.e., full modulation diversity since $n=2$), with a $b = \lceil 2\log_2(M) \rceil$-bit quantized receiver and minimum distance decoding given by (\ref{euc_dis_dec}) (See Appendix \ref{sec_Thm_full_div} and Theorem \ref{Thm_full_div}).
Subsequently, a rotated constellation which satisfies (\ref{matched_constel}) shall be referred as being {\em matched} to the quantizer.
It is easy to see that a rotated $M^2$-QAM constellation is matched to a $b=2\lceil \log_2(M) \rceil$-bit
uniform quantizer, if and only if, the projections of the transmit vectors are component-wise {\em equidistant}
and distinguishable.

Even with a mismatched rotated constellation having distinguishable projections (i.e., when the projections are not equidistant), full modulation
diversity may be achieved, but then the error probability would be higher, since some transmit vectors would be closer to the
edge of their quantization boxes (making it easier for noise
to move the transmitted vector to another quantization box when received) (illustrated through Fig.\ref{fig_non_equidistant} in Appendix \ref{sec_Thm_full_div}).
Following along the same lines as the proof in Theorem \ref{Thm_full_div}, it can be shown that {\em mismatched} constellations result in a higher error probability when compared to matched constellations.
This therefore leads us to the third code construction criterion.

{\em {\bf Criterion III} : In order to minimize the error probability of a rotated $M^2$-QAM constellation with a
$b= \lceil 2\log_2(M) \rceil$-bit quantized receiver, the rotation angle must be such that the rotated
$M^2$-QAM constellation is matched to the quantizer.}

\section {Rotated Constellation Design for Quantized Receiver}
\label{code_design}
In this section, we construct rotated $M^2$-QAM constellations which satisfy {Criterion III}.
We had earlier observed that, for $M^2$-QAM, a rotation by $\theta=\tan^{-1}(1/M)$ appeared
to be always in the set of {admissible} angles.
In fact, it can be shown analytically that a rotation by $\theta=\tan^{-1}(1/M)$,
satisfies Criterion III (See Theorem \ref{Thm_equi_code}, Appendix \ref{sec_Thm_equi_code}).

For $M^2$-QAM with $\theta=\tan^{-1}(1/M)$, it can be shown that the minimum product distance of the code is $4M/(M^2 + 1)$ ($\approx 4/M$ for $M \gg 1$).
On the other hand, a rotation angle of $\theta = 1/2 \tan^{-1}(2)$ is known to have a
minimum product distance of at least $4/\sqrt{5}$ irrespective of the QAM size.
Also, for any rotation angle the error performance with a quantized receiver is
inferior to that with an unquantized receiver.
Hence, with increasing $M$, the error performance of a {\em quantized} receiver with $\theta = \tan^{-1}(1/M)$ is expected to be increasingly less power efficient than that of a {\em unquantized} receiver with $\theta = 1/2 \tan^{-1}(2)$.
With increasing $M$, the set of
{admissible} angles appeared to be only
$\tan^{-1}(1/M) \pm \epsilon$ and therefore, it can be argued that,
the best possible error performance with a $b = \lceil 2\log_2(M) \rceil$-bit quantized receiver would have a loss
in power efficiency when compared to an unquantized receiver.
However, this appears to be the cost to achieve full modulation diversity in quantized receivers with limited precision.

\section { Imperfect Receiver knowledge of $\rho$ }
\label{imperfect_rho}
In the previous sections,
in order to achieve full modulation coding diversity, minimum distance decoding at the receiver assumed perfect knowledge of $\rho$.
In this section, we relax this assumption and present novel techniques
to estimate $\rho$ accurately.
It is expected that the error performance would degrade with imperfect
receiver knowledge of $\rho$. Interestingly, in sub-section \ref{rho_estimate_optimal} we propose an optimality criterion, which if satisfied by the estimate of $\rho$, would guarantee
{\em no loss} in the error probability performance of the minimum distance decoder with estimated $\rho$ when compared to the error performance with perfect knowledge of $\rho$.
Such an estimate would be referred to as an {\em optimal} estimate of $\rho$. 
We estimate $\rho$ based on the quantized receiver outputs for a known
transmitted sequence. We refer to this transmitted sequence as the $\rho$-training sequence.
Any $\rho$-training sequence which results in an optimal estimate of $\rho$ is subsequently referred
to as an optimal $\rho$-training sequence.
In sub-section \ref{rho_estimate_training} we present receiver control techniques required to estimate $\rho$.
Maximum likelihood estimation of $\rho$ based on the quantized receiver outputs of the $\rho$-training
sequence is discussed in sub-section \ref{sec_rho_estimate}.

Finally, in sub-section \ref{tr_seq_design}, for $M=2$ (rotated 4-QAM) we present an optimal $\rho$-training sequence
which satisfies the optimality criterion introduced in sub-section \ref{rho_estimate_optimal}.
For $M > 2$, the length of $\rho$-training sequences which satisfy the optimality criterion
is expected to be large resulting in too much training overhead and hence loss
in effective throughput. Therefore, a novel design of short $\rho$-training sequences is proposed,
which can achieve an error probability performance close to that achieved with optimal $\rho$-training sequences. Such short $\rho$-training sequences have been referred to as
`good' training sequences. Also throughout this section, it is assumed that {\em i}) with rotated $M^2$-QAM, a $b=\lceil 2\log_2(M) \rceil$-bit uniform quantizer is employed, {\em ii})
a minimum distance decoder is used for detection, and {\em iii}) the rotated constellation
satisfies Criterion III.
\subsection {Criterion for the optimal $\rho$ estimate} \label{rho_estimate_optimal}
Since the rotation matrix ${\bf G}$ is real-valued, it is obvious that the minimum distance decoder
in (\ref{euc_dis_dec}) separates into independent and identical minimum distance decoders for the real and imaginary components of the
transmitted information symbol vectors, and therefore
the error probability performance for both the real and imaginary components are also identical.
Hence, we only analyze the optimality of the minimum distance decoder with imperfect
$\rho$ estimate, only for the $I$ component.
The minimum distance decoder with the estimated $\rho$, is also given by
(\ref{euc_dis_dec}), but with $\rho$ replaced by its estimate ${\hat \rho}$.

We are now interested in studying the {\em conditions} under which the
error probability performance with ${\hat \rho}$ is exactly the same as the
error probability performance assuming perfect receiver knowledge of $\rho$.
Any estimate of $\rho$, which satisfies these conditions would be an {\em optimal
estimate} in terms of achieving an error probability performance same as that
achieved with perfect receiver knowledge of $\rho$.
To simplify notations, for any received vector ${\bf r}$, information symbol vectors ${\bf u}$ and ${\bf v}$
and any real $\zeta > 0$, we define
\begin{equation}
\label{E_dist_def}
m(\zeta,{\bf r}^I,{\bf u}^I) \Define {\Bigg (} {\bf r}^I - {\Big (} \frac{{\bf G}{\bf u}^I}{X} {\Big )} {\Bigg )}^{T} {\bf D}_{\zeta} {\Bigg (} {\bf r}^I - {\Big (} \frac{{\bf G}{\bf u}^I}{X} {\Big )} {\Bigg )}
\end{equation}
\vspace{-3mm}
\begin{equation}
\label{D_dist_def}
D_{E}(\zeta,{\bf r}^I,{\bf u}^I,{\bf v}^I) \Define {\Bigg (} m(\zeta,{\bf r}^I,{\bf u}^I) - m(\zeta,{\bf r}^I,{\bf v}^I) {\Bigg )}.
\end{equation}
The detected information symbols can therefore be stated in terms of $m(.)$ as
\begin{eqnarray}
\label{euc_dis_dec_IQ_1}
{\widehat {\bf u}}^I &=& \arg \min_{{\bf u}^I \in {\mathcal S}_M^2 }  m(\rho,{\bf r}^I,{\bf u}^I).
\end{eqnarray}
This then implies that, for any information symbol vector ${\bf v}$
\begin{equation}
\label {thm3_mot_1}
D_{E}(\rho,{\bf r}^I,{\widehat {\bf u}}^I ,{\bf v}^I) \leq 0
\end{equation}

With an estimated ${\hat \rho}$, if for all information symbol vectors ${\bf v} \in {\mathcal S}_M^2$
\begin{equation}
\label {thm3_mot_2}
D_{E}({\hat \rho},{\bf r}^I,{\widehat {\bf u}}^I ,{\bf v}^I) \leq 0
\end{equation}
then it is obvious that the output of the minimum distance decoder
with estimated $\rho$ is the {\em same} as the output of the minimum distance decoder
with perfect knowledge of $\rho$.
If (\ref{thm3_mot_2}) holds for all information symbol vectors ${\bf v} \in {\mathcal S}_M^2$, then along with
(\ref{thm3_mot_1}), it follows that
\begin{equation}
\label {thm3_mot_3}
D_{E}(\rho,{\bf r}^I,{\widehat {\bf u}}^I ,{\bf v}^I) D_{E}({\hat \rho},{\bf r}^I,{\widehat {\bf u}}^I ,{\bf v}^I) \geq 0
\end{equation}
for all information symbol vectors ${\bf v} \in {\mathcal S}_M^2$.
Since ${\widehat {\bf u}}^I$ could be any information symbol vector in ${\mathcal S}_M^2$
and $D_{E}(\zeta,{\bf r}^I,{\bf u}^I,{\bf v}^I) = -D_{E}(\zeta,{\bf r}^I,{\bf v}^I,{\bf u}^I)$,
it is easy to see that the output of the minimum distance decoder with estimated ${\hat \rho}$
would be the same as that with perfect knowledge of $\rho$ if
\begin{eqnarray}
\label{thm1_statement}
D_{E}(\rho,{\bf r}^I,{\bf u}^I,{\bf v}^I) D_{E}({\hat \rho},{\bf r}^I,{\bf u}^I,{\bf v}^I) \geq 0
\end{eqnarray}
for all possible received vector ${\bf r}$ (finitely many due to receiver quantization) and all possible information symbol vectors ${\bf u}$ and ${\bf v}$.
We formally prove this observation in the following theorem.

\begin{mytheorem}\label{Thm1}
For a given realization of $\rho$,
and estimated ${\hat \rho}$, if (\ref{thm1_statement}) is satisfied
for all possible received vector ${\bf r}$ and all possible information symbol vectors ${\bf u}$ and ${\bf v}$,
then ${\hat \rho}$ is an {\em optimal} estimate of $\rho$.
\end{mytheorem}

{\em Proof}:
See Appendix \ref{sec_Thm1}.
$\hfill \blacksquare$

We now analyze the condition set-forth in Theorem \ref{Thm1} regarding the {\em optimal}
estimate of ${\rho}$.
With each information symbol belonging to $M^2$-QAM, and $b = \lceil 2 \log_2(M) \rceil$
we make the following definitions
\begin{eqnarray}
\label{theorem2_defs}
{\mathcal D}_M   \Define  {\Big \{} \frac{(a_1 - a_2)}{2^b - 1} \,|\, a_1,a_2 \in {\mathcal S}_{M^2} {\Big \}} & , & {\mathcal D}_M^2  \Define  {\Big \{} a^2 \,|\, a \in {\mathcal D}_M {\Big \}} \nonumber \\
{\mathcal Q}_M  \Define  {\Big \{} \frac{(a_1 - a_2)}{(a_3 - a_4)} \,|\, a_1,a_2,a_3,a_4 \in {\mathcal D}_M^2\,,\, a_3 \ne a_4 {\Big \}} & , & {\mathcal Q}_M^{+}  \Define  {\Big \{} a \,|\, a \in {\mathcal Q}_M \, , \, a \geq 0 {\Big \}}
\end{eqnarray}
where ${\mathcal S}_{M^2}$ is the $M^2$-PAM signal set (see Section \ref{sec2} for PAM set definition).
It is also noted that ${\mathcal S}_{M^2}$ is not the same as ${\mathcal S}_M^2$.
As an example, with $M=2$ and $b = \lceil 2 \log_2(M) \rceil = 2$, ${\mathcal S}_{M^2} = \{ -3, -1, 1, 3 \}$ and ${\mathcal D}_M = \{ -2, -4/3, -2/3, 0, 2/3, 4/3, 2 \}$.
Based on Theorem \ref{Thm1}, the next theorem gives a useful sufficiency condition for the optimality of
an estimate of $\rho$.

\begin{mytheorem}\label{Thm2}
Consider a rotated $M^2$-QAM constellation matched to the quantizer. Let ${\hat \rho}$ be an estimate of ${\rho}$ satisfying the following condition
\begin{eqnarray}
\label{thm2_statement}
\forall \, l \in {\mathcal Q}_M^{+} & : & \rho^2 \leq l \,\Rightarrow\, {\hat \rho}^2 \leq l \nonumber \\
\forall \, l \in {\mathcal Q}_M^{+} & : & \rho^2 \geq l \,\Rightarrow\, {\hat \rho}^2 \geq l 
\end{eqnarray}
Then, ${\hat \rho}$ is an {\em optimal} estimate of $\rho$.
\end{mytheorem}

{\em Proof} : See Appendix \ref{sec_Thm2}.
$\hfill \blacksquare$

Let ${\mathcal Q}_M^{+} = \{ 0, q_1, q_2, \cdots ,q_{L_M}\}$, with $0 < q_1 < q_2 < \cdots < q_{L_M}$.
It is clear that the elements of the set ${\mathcal Q}_M^{+}$ {\em partition} the positive real line $[0 , \infty)$
into $(L_M + 1)$ intervals with $[0 , q_1)$ and
$[q_{L_M} , \infty)$ being the first and the last interval respectively. The $k$-th intermediate interval is given by $[q_{k-1} , q_{k})$, $k=2,3, \cdots ,L_M$.
For any finite set ${\mathcal S} = \{ s_1, s_2, \cdots , s_n \}$
with $0 \leq s_1 < s_2 < \cdots < s_n$, let ${\mathcal I}({\mathcal S})$ be the set of intervals {\em induced}
by the set ${\mathcal S}$. That is
\begin{equation}
\label{def_I_S}
{\mathcal I}({\mathcal S}) = {\Big \{}  [ s_1 , s_2 ) \,,\,  [ s_2 , s_3 ) \,,\, \cdots \,,\, [ s_n , \infty ) {\Big \}}
\end{equation}

The sufficiency condition in (\ref{thm2_statement}) can now be understood in terms of the $(L_M + 1)$ intervals
of the positive real line induced by the set ${\mathcal Q}_M^{+}$.
The sufficiency condition basically states that, for an estimate of $\rho$ to be
{\em optimal}, it {\em must} belong to the same interval of ${\mathcal I}({\mathcal Q}_M^{+})$ in which $\rho$ lies.
This therefore also implies that, for an estimate ${\hat \rho}$ to be {\em optimal}
it is {\em not} necessary that ${\hat \rho}$ be exactly equal to ${\rho}$.

\subsection {Receiver control for estimating $\rho$}\label{rho_estimate_training}
In this section, we discuss receiver control techniques required for estimating ${\rho}$. 
In the proposed rotation coding scheme, coding is performed across $n=2$ channel realizations.
Using a 2-dimensional rotation matrix ${\bf G}$, a pair of information symbols
is transformed into a pair of coded output symbols. The first coded symbol in the pair
is transmitted during channel realization 1 (with channel gain $\vert h_1 \vert$), whereas the second coded symbol
is transmitted during channel realization 2 (with channel gain $\vert h_2 \vert$).
For both channel realizations, the preamble sequence used for
tuning the VGA gain is the same as discussed in Section \ref{sec2}. However, for channel realization 2, even before transmitting
this preamble sequence, a known $\rho$-training sequence is transmitted for estimating $\rho$. During the transmission of the
$\rho$-training sequence the analog gains $g_{AFE}$ and $g_{VGA}$ are set to the values
programmed during the transmission of coded information symbols in channel realization 1, and therefore
\begin{equation}
\label{gains_first_cmp}
g_{VGA} g_{AFE} = \frac {1}{\vert h_1 \vert X}.
\end{equation}

The $\rho$-training sequence
is a sequence of $l$ distinct {\em positive} valued symbols with each symbol
being transmitted multiple times to average out the effect of receiver noise
\footnote{\footnotesize {At the receiver, depending upon the repetition factor of the $\rho$-training symbols, the cut-off frequency of the LPF is appropriately reduced, which helps in noise reduction.}}.
Subsequently, we shall denote an arbitrary $\rho$-training sequence by ${\mathcal T}$.
Let the $k$-th training symbol be given by $c_k$, $k=1,2,\cdots,l$.
The $l$ corresponding inputs to the sample and hold
circuit are given by
\begin{equation}
\label{sh_inp_training}
s_k = g_{AFE} g_{VGA} \vert h_2 \vert c_k \,,\, k=1,2, \cdots l
\end{equation}
Using (\ref{gains_first_cmp}) in (\ref{sh_inp_training}), the sample and hold,
and quantizer outputs during the transmission of the $\rho$-training sequence in channel realization 2 are given by
\begin{eqnarray}
\label{adc_out_training}
s_k  \, = \, \frac {\vert h_2 \vert c_k}{\vert h_1 \vert X}  \, = \, \rho \frac{c_k}{X}\,\,,\,\,k=1,2,\ldots,l \\
r_k = Q_b {\Big (} s_k {\Big )} = Q_b {\Big (} \rho \frac{c_k}{X} {\Big )} \,,\, k=1,2, \ldots l.
\end{eqnarray}
In channel realization 2, after all the $\rho$-training symbols are transmitted,
the receiver estimates $\rho$ based on the $l$ observations $\{ r_k , k=1,2\cdots,l \}$.

\subsection {Estimation of $\rho$}
\label{sec_rho_estimate}
The $l$ discrete outputs of the ADC ($\{ r_1, r_2, \cdots r_l\}$) can be used to estimate $\rho$ as follows.
Given the $l$ discrete outputs, the maximum likelihood estimate (MLE) of $\rho$ is given by
\begin{equation}
\label{ml_estimate_rho}
\rho_{ML} = \arg \max_{\rho > 0} P( r_1, r_2, \cdots r_l | \rho , \{ c_k \} , l)
\end{equation}
where $P(r_1, r_2, \cdots r_l | \rho , \{ c_k \} , l)$ is the probability that the $l$ outputs
take the values $\{ r_1, r_2, \cdots r_l\}$ for a given channel gain ratio $\rho$, and
the $\rho$-training sequence $\{ c_k \}$.
For the $k$-th training symbol $c_k$, since
$r_k = Q_b(\rho c_k / X)$, from (\ref{qri}) it must be true that
\begin{eqnarray}
\label{segment_k}
\frac {r_k - \frac{1}{2^b - 1}} {c_k / X } \leq  \rho  <  \frac {r_k + \frac{1}{2^b - 1}} {c_k / X}\,\,,\,\, \mbox{if} \,\, r_k < 1 \\ \nonumber
\frac {1 - \frac{1}{2^b - 1}} {c_k / X } \leq  \rho  <  \infty \,\,,\,\, \mbox{if} \,\, r_k = 1
\end{eqnarray}
The inequality in (\ref{segment_k}) defines an interval of the positive real line, which we shall denote by ${\mathcal L}_k\,\,,\,\,k=1,2,\ldots,l$. Therefore the $l$ outputs would be
$\{ r_1, r_2, \cdots r_l\}$ if and only if $\rho \in {\mathcal L}$, where
${\mathcal L} \Define \cap_{k=1}^{l} {\mathcal L}_k$. Further, we would call ${\mathcal L}$ as the ``ML interval''
corresponding to the training sequence $\{ c_k \}$ and the $l$ outputs $\{ r_1, r_2, \cdots r_l\}$.
Also, given the $l$ outputs, all the values of $\rho$ in the interval
${\mathcal L}$ are equally probable. Let ${\mathcal L}_{sup} \Define \mbox{sup} \, {\mathcal L}$, and ${\mathcal L}_{inf} \Define \mbox{inf} \,{\mathcal L}$, denote the supremum and infimum of the
interval ${\mathcal L}$.
One possible ML estimate of $\rho$, that we propose,
is then given by
\begin{equation}
\label{rho_hat}
{\widehat \rho} = \left\{\begin{array}{cc}
\frac { {\mathcal L}_{sup} + {\mathcal L}_{inf}  }{2} &  {\mathcal L}_{sup} < \infty\\
{\mathcal L}_{inf} & \mbox{otherwise}.
\end{array} \right.
\end{equation}
For a given $\rho$-training sequence ${\mathcal T} = \{ c_1, c_2, \cdots, c_l  \}$, and a set of corresponding outputs
${\mathcal R} = \{ r_1, r_2, \cdots r_l \}$, let ${\mathcal L}({\mathcal T},{\mathcal R}) \subset {\mathbb R}^{+}$ denote the ML interval.
 
As an example, let us consider a $b$=$2$-bit quantizer, and a training sequence $\{ c_k \} = \{X/4, X/2, X, 2X, 4X \}$.
Let the $l=5$ corresponding output symbols of the quantizer be $\{ r_1=1/3, r_2=1/3, r_3=1 , r_4=1 , r_5=1 \}$. The intervals ${\mathcal L}_k$ corresponding to these $5$ outputs are
\begin{eqnarray}
\label{ex_lineseg}
{\mathcal L}_1 : \, 0 \leq \frac{\rho}{4} < \frac{2}{3} \,,\,
{\mathcal L}_2 : \, 0 \leq \frac{\rho}{2} < \frac{2}{3} \,,\,
{\mathcal L}_3 : \, \frac{2}{3} \leq {\rho} < \infty \,,\,
{\mathcal L}_4 : \, \frac{2}{3} \leq {2\rho} < \infty \,,\,
{\mathcal L}_5 : \, \frac{2}{3} \leq {4\rho} < \infty.
\end{eqnarray}
The ML interval ${\mathcal L}(\{X/4, X/2, X, 2X, 4X \}, \{1/3,1/3,1,1,1 \} ) = [2/3 \,,\, 4/3 )$ and hence ${\widehat \rho} = 1$.

For a $b$-bit quantizer and some fixed $\rho$-training sequence ${\mathcal T} = \{ c_1, c_2, \cdots , c_l \}$ of $l$ training symbols,
it is clear that for each value of $\rho \in [0 , \infty)$,
there is a corresponding output sequence ${\mathcal R}(\rho, {\mathcal T}) = \{ r_k = Q_b(\rho \frac{c_k}{X})\,,\,k=1,2,\cdots,l\}$. We shall refer to each such possible output sequence
as a {\em feasible} output sequence for the given training sequence. Note that even though the range of values of $\rho$ is infinite, the number of distinct feasible output sequences is finite due to the finite length of the $\rho$-training sequence and the finite number ($2^b$) of quantizer levels for a $b$-bit uniform quantizer. Further, for each feasible
output sequence ${\mathcal R}'$, there exists an ML interval ${\mathcal L}({\mathcal T}, {\mathcal R}')$. 
Also, it is trivially true that
two ML intervals corresponding to two different feasible output sequences are disjoint (i.e., if ${\mathcal R}' \ne {\mathcal R}''$ then ${\mathcal L}({\mathcal T}, {\mathcal R}') \cap {\mathcal L}({\mathcal T}, {\mathcal R}'') = \phi$, where $\phi$ denotes the null set).
In addition, since $\rho \in [0 , \infty)$, it follows that the ML intervals corresponding
to all possible feasible output sequences, form a {\em partition} of the positive real line.
This is summarized as follows.
For any $0 \leq \rho_1,\rho_2 < \infty \,,\,\rho_1 \ne \rho_2 $
\begin{eqnarray}
\label{ml_line_interv}
{\mathcal L}({\mathcal T}, {\mathcal R}(\rho_1, {\mathcal T})) \cap {\mathcal L}({\mathcal T}, {\mathcal R}(\rho_2, {\mathcal T}))  = \left\{\begin{array}{cc}
\phi &, \,\, {\mathcal R}(\rho_1, {\mathcal T}) \ne {\mathcal R}(\rho_2, {\mathcal T}) \\
{\mathcal L}({\mathcal T}, {\mathcal R}(\rho_1, {\mathcal T})) = {\mathcal L}({\mathcal T}, {\mathcal R}(\rho_2, {\mathcal T})) &, \,\, {\mathcal R}(\rho_1, {\mathcal T}) = {\mathcal R}(\rho_2, {\mathcal T})
\end{array} \right.
\end{eqnarray}
Also,
\begin{equation}
\bigcup_{_{_{\rho \in [0 , \infty)}}}  \hspace{-2mm} {\mathcal L}({\mathcal T}, {\mathcal R}(\rho, {\mathcal T}))  =  {\mathbb R}^{+}.
\end{equation}
Some more interesting properties are as follows.
For any $\rho$-training sequence ${\mathcal T}$, we have the following interesting properties.
\begin{eqnarray}
\label {rho_stmts}
\rho & \in & {\mathcal L}({\mathcal T}, {\mathcal R}(\rho, {\mathcal T}))\,,\, \,\, 0 \leq \rho < \infty \nonumber \\
\rho_1 \in {\mathcal L}({\mathcal T}, {\mathcal R}(\rho_2, {\mathcal T})) & \Rightarrow  & \rho_2 \in {\mathcal L}({\mathcal T}, {\mathcal R}(\rho_1, {\mathcal T})) \,\, \mbox{and vice-versa} \,\, \,\, 0 \leq \rho_1,\rho_2 < \infty \,,\,\rho_1 \ne \rho_2 \nonumber \\
{\hat \rho} & \in & {\mathcal L}({\mathcal T}, {\mathcal R}(\rho, {\mathcal T}))\,,\, \,\, 0 \leq \rho < \infty
\end{eqnarray}
where ${\hat \rho}$ is the ML estimate of $\rho$ given by (\ref{rho_hat}).
For any ${\mathcal T}$, $\rho' \in {\mathcal L}({\mathcal T}, {\mathcal R}(\rho, {\mathcal T}))$, if and only if
${\mathcal R}(\rho, {\mathcal T})) = {\mathcal R}(\rho', {\mathcal T}))$. This equality is trivially
satisfied for $\rho' = \rho$, which completes the proof for the first statement in (\ref{rho_stmts}).
For the second statement, we note that $\rho_1 \in {\mathcal L}({\mathcal T}, {\mathcal R}(\rho_2, {\mathcal T}))$ if and only if ${\mathcal R}(\rho_1, {\mathcal T})) = {\mathcal R}(\rho_2, {\mathcal T}))$.
This then implies that $\rho_2 \in {\mathcal L}({\mathcal T}, {\mathcal R}(\rho_1, {\mathcal T}))$.
The third statement follows from the fact that, for any $\rho$, the ML estimate ${\hat \rho}$ is always between the supremum and the infimum of the interval ${\mathcal L}({\mathcal T}, {\mathcal R}(\rho, {\mathcal T}))$.  

Given a fixed $\rho$-training sequence ${\mathcal T}$, from (\ref{rho_stmts}) it follows that both $\rho$ and the proposed ML estimate ${\hat \rho}$ lie in the ML interval ${\mathcal L}({\mathcal T}, {\mathcal R}(\rho, {\mathcal T}))$.
In addition to this, if the ML interval corresponding to each feasible output sequence
is a subset of some interval induced by the set ${\mathcal Q}_M^{+}$, then
for any $l \in {\mathcal Q}_M^{+}$ it follows that if $l$ is greater than $\rho$ then it is also greater than
${\hat \rho}$, and similarly when $l$ is smaller than $\rho$ then it is also smaller than ${\hat \rho}$.  
However this is precisely the sufficiency condition in Theorem \ref{Thm2}.
We can therefore conclude that
with the proposed $\rho$ estimation technique (see (\ref{rho_hat})), a training sequence ${\mathcal T}$ results in an optimal estimate of ${\rho}$, if ${\mathcal T}$ satisfies the following conditions.
\begin{eqnarray}
\label {rho_hat_optimality}
\forall \,\, 0 \leq \rho < \infty & : & {\mathcal L}({\mathcal T}, {\mathcal R}(\rho,{\mathcal T})) \subseteq {I}\, , \mbox{for some} \, {I} \in {\mathcal I}({\mathcal Q}_M^{+}). 
\end{eqnarray}
We next design optimal and near-optimal $\rho$-training sequences based on the criterion in (\ref{rho_hat_optimality}).
\subsection {Design of $\rho$-training sequence for estimating $\rho$} \label{tr_seq_design}
We first show that with $M=2$ and a $b=2$-bit uniform quantizer it is possible to design a $\rho$-training
sequence which satisfies (\ref{rho_hat_optimality}) and is therefore optimal. 

\begin{mytheorem}\label{Thm3}
Consider a rotated constellation matched to the quantizer. Let $M=2$, $b=2$ and ${\mathcal Q}_M^{+} = \{ 0, q_1, q_2, \cdots
q_{L_M}\}, 0 < q_1 < q_2 < \cdots < q_{L_M}$. The following $\rho$-training sequence $\{ c_k \}$ of length $L_M$ with the proposed ML estimator (Section \ref{sec_rho_estimate}) results in an {\em optimal}
estimate of $\rho$.
\begin{eqnarray}
\label{thm3_statement}
c_k = \frac{2}{3} \frac{X}{q_{_{L_{_M} - k + 1}}}~~,~~k=1,2,\cdots,L_M.
\end{eqnarray}
\end{mytheorem}

{\em Proof}: See Appendix \ref{sec_Thm3}.
$\hfill \blacksquare$

For a general $M > 2$, 
it is challenging to design an optimal $\rho$-training sequence
which satisfies the sufficiency condition in Theorem \ref{Thm2}.
Further,
we conjecture that, just as with $M=2$,
for any $M > 2$ also, the length of optimal
$\rho$-training sequences based on Theorem \ref{Thm2} would be proportional to $L_M=\vert {\mathcal Q}_M^{+} \vert$.
However, the cardinality of ${\mathcal Q}_M^{+}$ is
a rapidly increasing function of $M$ (e.g., $\vert {\mathcal Q}_2^{+} \vert= 29$, and $\vert {\mathcal Q}_4^{+} \vert=3939$), which then implies that
with increasing $M$ a significant amount of communication bandwidth would be used up in the transmission of the $\rho$-training
sequence, resulting in reduced overall throughput. Therefore, it is of {\em practical} interest to design $\rho$-training sequences which are short
and which can still achieve an error performance comparable to that achieved with optimal $\rho$-training sequences.

Towards designing such {\em practical} sequences, we observe that the average error performance
with estimated $\rho$, would be sensitive to the amount of overlap between the ML intervals induced\footnote {\footnotesize {These are basically the ML intervals corresponding to all feasible output sequences for the given $\rho$-training sequence.}} by the
$\rho$-training sequence and the intervals induced by ${\mathcal Q}_M^{+}$.
With short $\rho$-training sequences, the
ML intervals induced by the $\rho$-training sequence would not coincide exactly with the intervals induced
by ${\mathcal Q}_M^{+}$. Nevertheless, it may be possible to
design short length $\rho$-training sequences for which some of the ML intervals belong to ${\mathcal I}({\mathcal Q}_M^{+})$. With short $\rho$-training sequences, an interval induced by
${\mathcal Q}_M^{+}$, which is exactly the same as some ML interval induced by the $\rho$-training sequence, shall be referred to as ``covered'' by that $\rho$-training sequence.
\begin{figure}[t]
\begin{center}
\epsfig{file=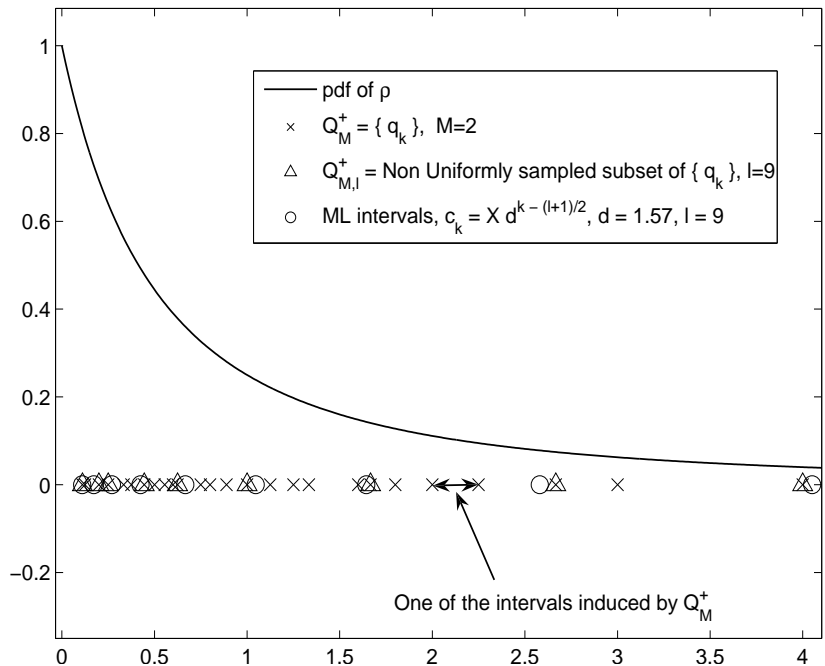, width=110mm,height=90mm}
\end{center}
\vspace{-5mm}
\caption{The intervals induced by the set ${\mathcal Q}_M^+$ for $M=2$. ML intervals for the proposed
$\rho$-training and estimation schemes.}
\vspace{-5mm}
\label{pdf_rho_line_seg}
\end{figure}

From Fig.~\ref{pdf_rho_line_seg}, we try to gain more insights into the problem of designing
shorter length $\rho$-training sequences for $M=2$.
We observe that the density of the intervals induced by ${\mathcal Q}_M^{+}$ (depicted with cross `X' marks
on the horizontal axis) is much more higher
near the origin than farther away.
Furthermore, from the p.d.f. of $\rho = \vert h_2\vert /\vert h_1 \vert$ (Rayleigh faded $h_1$ and $h_2$), we observe that most of the probability
mass is distributed near the origin\footnote {\footnotesize {In fact, for any other fading distribution also,
it can be shown that $P(\rho < 1) = P(\rho > 1) = 1/2$.}}. Based on these observations and the sufficiency conditions in Theorem \ref{Thm2}, it can be argued
that, to have an error performance comparable to that of an optimal $\rho$-training sequence, any short $\rho$-training sequence should aim to ``cover'' the intervals of ${\mathcal I}({\mathcal Q}_M^{+})$
which are closer to the origin. This reasoning is supported by two facts.
Firstly, with i.i.d. channels gains $\vert h_1 \vert$ and $\vert h_2 \vert$, the
probability of $\rho$ taking large values is small, and hence large values of $\rho$ are expected
to have lesser contribution to the average error probability than smaller values of $\rho$.
Secondly, when $\rho >> 1$,
any error in the estimation of $\rho$ is likely to have a lesser impact on the error performance compared
to when $\rho < 1$.
To see this, we note that for $\rho >> 1$, the ML estimate for any $\rho$-training sequence would be the infimum value
of the ML interval corresponding to the all ones output sequence
which would also be large i.e., ${\hat \rho} >> 1$.
Therefore for any two transmit vectors ${\bf x} = (x_1 , x_2)^T = {\bf G}{\bf u}$ and ${\bf y} = (y_1 , y_2)^T = {\bf G}{\bf v}$, $D_{E}(\rho,{\bf r}^I,{\bf u}^I,{\bf v}^I) D_{E}({\hat \rho},{\bf r}^I,{\bf u}^I,{\bf v}^I) \approx \rho^2 {\hat \rho}^2 d_2^2 > 0$, where $d_2 = (r_2^I - x_2^I/X)^2 - (r_2^I - y_2^I/X)^2$.
Using Theorem \ref{Thm1}, this then implies that, with high probability, the output of the minimum distance decoder with estimated $\rho$ is the same as its output
with perfect knowledge of $\rho$.

Therefore, any short $\rho$-training sequence should aim to ``cover'' the intervals of ${\mathcal I}({\mathcal Q}_M^{+})$
which are closer to the origin.
With $M=2$, in Fig.~\ref{pdf_rho_line_seg}
a short $\rho$-training sequence of length $l < L_M$ is designed
in a way to ``cover'' only the intervals of ${\mathcal I}({\mathcal Q}_M^{+})$ which are closer to origin.
This is done by {\em non-uniformly} sampling\footnote{\footnotesize {we use the word ``non-uniform'' since the sampling is biased towards choosing more elements which are closer to the origin.}} out $l$ distinct elements of the set ${\mathcal Q}_M^{+}$, such that
the intervals induced by these $l$ elements coincide with {\em most} of the intervals of
${\mathcal I}({\mathcal Q}_M^{+})$ which are near to origin. Let us denote this set of $l$ elements as
${\mathcal Q}_{(M,l)}^{+}$. The corresponding short $\rho$-training sequence which has ML intervals
coinciding exactly with the intervals induced by ${\mathcal Q}_{(M,l)}^{+}$
is then given by (\ref{thm3_statement}), where the set ${\mathcal Q}_M^{+}$ is replaced by the set
${\mathcal Q}_{(M,l)}^{+}$ and $L_M$ is replaced by $l$.
In Fig.~\ref{pdf_rho_line_seg}, for $l=9$, the elements of {\em one such} ${\mathcal Q}_{(M,l)}^{+}$ are
depicted through `triangles'.
Short $\rho$-training sequences which
achieve an error performance close to that achieved with optimal $\rho$-training sequences,
would be subsequently referred to as `good' $\rho$-training sequences.

Even though `non-uniform' sampling of ${\mathcal Q}_M^+$ is one possible
method for designing `good' $\rho$-training sequences,
with increasing $M$, the number of ways in which `non-uniform' sampling can be done, would
also increase rapidly, thereby increasing the complexity of finding `good'
$\rho$-training sequences.
Therefore for large $M$, a simpler strategy is required to search for `good'
$\rho$-training sequences. We next present a very simple and parameterizable
short $\rho$-training sequence design, which results in `good' $\rho$-training sequences. The $k$-th symbol of the proposed
$\rho$-training sequence is given by
\begin{equation}
\label{exp_tr_design}
c_k = X d^{(k - \frac{l+1}{2})}~~,~~k=1,2,\cdots,l
\end{equation}
where $l$ is the length of the training sequence, and
$d > 1$ is the ratio between the consecutive ${\rho}$-training symbols.
Let us denote this $\rho$-training sequence by ${\mathcal T}_d$.
This design is based on the observation that
for many `non-uniformly' sampled subsets ${\mathcal Q}_{(M,l)}^{+} \subset {\mathcal Q}_{M}^{+}$, it is possible
to find a value of $d$, such that a $\rho$-training sequence designed using (\ref{exp_tr_design}), would
have ML intervals ``almost'' same as the ML intervals of the $\rho$-training sequence designed using the `non-uniformly'
sampled subset. That is, for many non-uniformly sampled subsets ${\mathcal Q}_{(M,l)}^{+} \subset {\mathcal Q}_{M}^{+}$, for any $\rho > 0$, there exists some $d > 1$ and some $\epsilon$ close to zero, such that
\begin{equation}
\label {tr_design_cr}
\vert {\mathcal L}({\mathcal T}_d, {\mathcal R}(\rho, {\mathcal T}_d)) \, \cap \,  {\mathcal L}({\mathcal T}', {\mathcal R}(\rho, {\mathcal T}'))^c \vert \,\, \leq \,\, \epsilon \vert  {\mathcal L}({\mathcal T}', {\mathcal R}(\rho, {\mathcal T}'))  \vert
\end{equation}
where ${\mathcal T}'$ refers to the $\rho$-training sequence designed using the given
non-uniformly sampled subset ${\mathcal Q}_{(M,l)}^{+} \subset {\mathcal Q}_{M}^{+}$
\footnote{\footnotesize {For any real interval $I$, $\vert I \vert \Define (\mbox{sup} \, I  - \mbox{inf} \, I)$ refers to the length of the interval and $I^c$ refers to the complementary set ${\mathbb R} - I$ (i.e., all real numbers which do not belong to $I$).}}. 
For $M=2$, this fact is illustrated through Fig.~\ref{pdf_rho_line_seg}, where
for the given `non-uniformly' sampled subset of ${\mathcal Q}_M^{+}$, i.e., ${\mathcal Q}_{(M,l)}^{+}$ (shown with triangles),
a $\rho$-training sequence designed using (\ref{exp_tr_design}) has ML intervals (shown with circles) almost coinciding with the intervals induced by ${\mathcal Q}_{(M,l)}^{+}$.
For a given $M$ and $l$, the optimal
$d$ can be found at reasonable complexity, by minimizing the average error probability (as a function of $d$) using Monte-Carlo techniques.
\section {Simulation Results}
\label{sec_sim}
\begin{figure}[t]
\begin{center}
\hspace{-7mm}
\epsfig{file=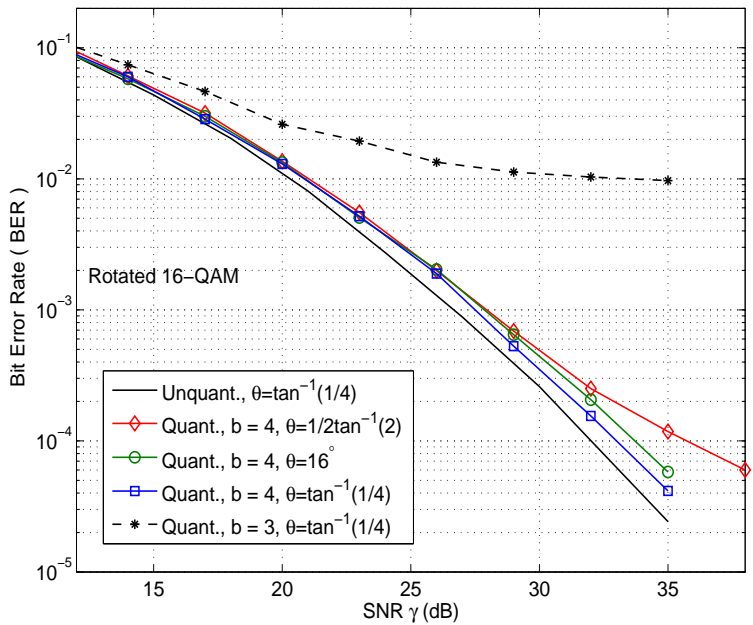, width=100mm,height=85mm}
\end{center}
\vspace{-4mm}
\caption{BER vs. SNR for a quantized receiver. $b=4$, 16-QAM and perfect channel state information at receiver.}
\vspace{-4mm}
\label{siso_16qam_b4}
\end{figure}
All error probabilities reported in this section have been averaged over the Rayleigh
flat fading statistics of the channel. Also, the receiver is assumed to have perfect channel state information. 
In Fig. \ref{siso_16qam_b4}, we plot the average bit error rate/probability (BER), for rotated
16-QAM constellation ($M$=$4$) and a $b$=$4$-bit quantized receiver. 
The following four important observations can be made in Fig. \ref{siso_16qam_b4}: {\em i}) with $\theta$ = $1/2 \tan^{-1}(2)$ (which is known to achieve full modulation diversity in an unquantized receiver, but does not satisfy Criterion II), the BER performance with a quantized receiver
fails to achieve full diversity (note the difference in slope at high SNR), which validates Criterion II, {\em ii}) with $\theta$ = $\tan^{-1}(1/4)$, which results in equidistant projections, the
quantized receiver achieves full modulation diversity with $b$=$4$. Further, the quantized receiver performs only 1 dB away from an ideal {\em unquantized} receiver at a BER of $10^{-4}$, {\em iii}) with a quantized receiver a rotation angle of $\theta$ = $16^{\circ}$ also appears to achieve full modulation diversity, but perform poor when compared to a matched rotated constellation with $\theta$ = $\tan^{-1}(1/4)$. This supports
Criterion III,
and {\em iv})
In Fig.~\ref{siso_16qam_b4} it is also
observed that with 16-QAM rotated constellation ($\theta$ = $\tan^{-1}(1/4)$), the error performance floors with $b$=$3 < 4$ quantization bits, which validates code design Criterion I.
\begin{figure}[t]
\begin{center}
\hspace{-7mm}
\epsfig{file=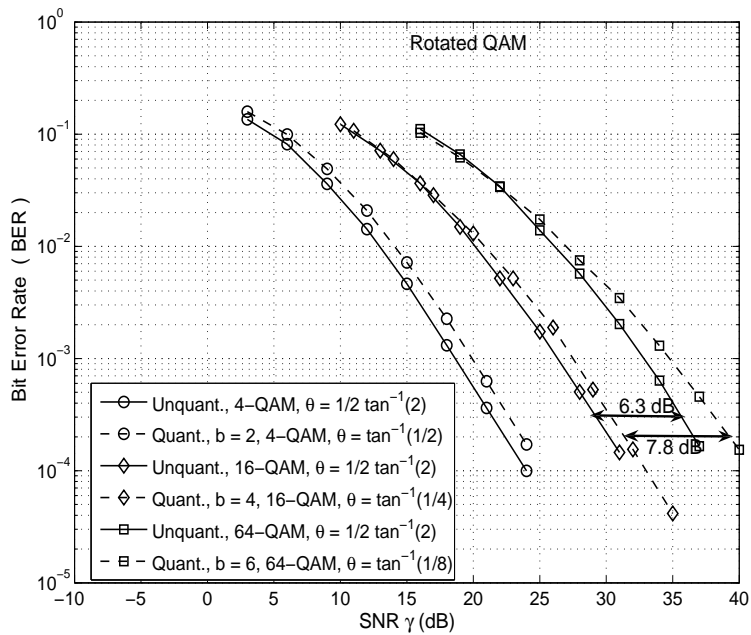, width=100mm,height=85mm}
\end{center}
\vspace{-5mm}
\caption{BER comparison between quantized and unquantized receivers. 4-,16-,64-QAM. Perfect
channel state information at receiver.}
\vspace{-4mm}
\label{siso_qam_blogM}
\end{figure}

It was discussed in Section \ref{code_design}, that
with increasing QAM size, a quantized receiver would be increasingly
less power efficient when compared to an unquantized receiver. This fact is illustrated in Fig. \ref{siso_qam_blogM}, where
the BER performance of both unquantized receiver with $\theta$=$1/2 \tan^{-1}(2)$ and quantized receiver
with $\theta$=$\tan^{-1}(1/M)$ are plotted for $M^2$=4-,16- and 64-QAM and $b$=$\lceil 2 \log_2(M) \rceil$.
Perfect channel state information is assumed at the receiver.
It is observed that for a fixed BER of $2 \times 10^{-4}$, with increasing QAM size, the increase in signal power required by a quantized receiver is more than that for an unquantized receiver.
An unquantized receiver requires $6.3$ dB more transmit power when the
QAM size is increased from 16 to 64. For the same increase in QAM size, a quantized
receiver would require $7.8$ dB more transmit power. However, when the QAM size is
increased from 4 to 16, the extra transmit power required for a fixed target BER
of $2 \times 10^{-4}$ is roughly the same (about $7.7$ dB) for both quantized and unquantized
receiver.
In Fig.\ref{siso_qam_crit3} (Appendix \ref{sec_Thm_full_div}), we report the BER performance of a rotated 16-QAM constellation for varying $\theta$ and fixed SNR.
It is observed that, the rotation angle $\theta = \tan^{-1}(1/M)$ (which results in a matched
constellation) achieves the minimum BER. This then supports code design Criterion III.
\begin{figure}[t]
\begin{center}
\hspace{-7mm}
\epsfig{file=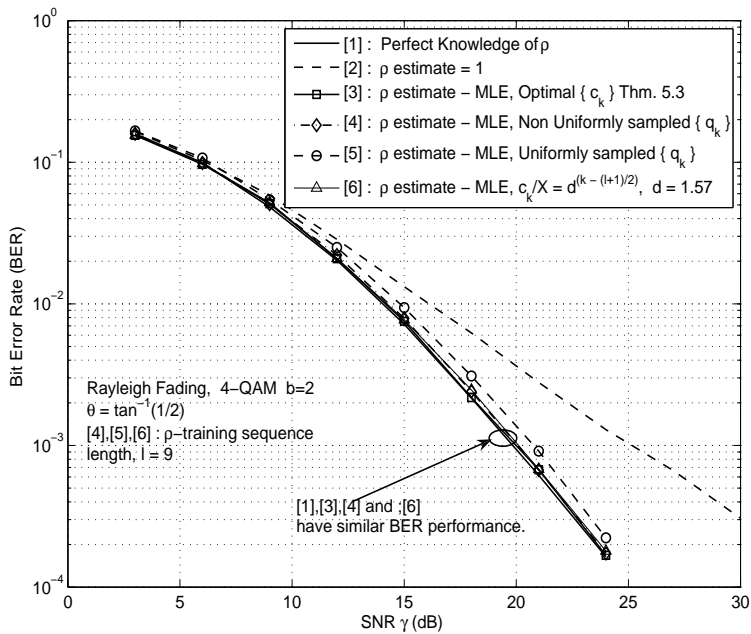, width=100mm,height=85mm}
\end{center}
\vspace{-5mm}
\caption{BER performance with a quantized receiver ($b=2$) and imperfect receiver knowledge of $\rho$. Rotated 4-QAM ($M=2$).}
\vspace{-4mm}
\label{rho_est_perf_b2_4qam}
\end{figure}

In Fig. \ref{rho_est_perf_b2_4qam}, the BER performance with minimum distance decoding and imperfect knowledge of $\rho$, is plotted as a function of $\gamma$ for $M=2$. The rotation angle
is $\theta = \tan^{-1}(1/2)$ and $b=2$.
We firstly make a note that, for a matched rotated constellation, the error probability performance with ${\hat {\rho}}$ being any arbitrary positive valued estimate of ${\rho}$, {\em does not have error
floors}.
This is because, in the absence of noise (i.e., $\gamma = \infty$), when a certain information symbol vector ${\bf v}$ is transmitted, with ${\bf r}$ as the quantized output vector, the detection metric of some information symbol vector ${\bf u}$
(i.e., $m({\hat \rho},{\bf r}^I,{\bf u}^I)$ and $m({\hat \rho},{\bf r}^Q,{\bf u}^Q)$ )
is equal to zero only for ${\bf u} = {\bf v}$, and is positive for all other
possible information symbol vectors.
Therefore, the detected information symbol vector ${\widehat {\bf u}}$
is the same as the transmitted vector ${\bf v}$, resulting in zero probability of error.
This argument is supported by the fact that in Fig.~\ref{rho_est_perf_b2_4qam}, a fixed estimate of ${\hat \rho}=1$, has no
floors in its BER performance.

In Fig.~\ref{rho_est_perf_b2_4qam} we also observe that the BER performance of the {\em optimal} $\rho$-training sequence
designed using Theorem \ref{Thm3} is same as the BER achieved with perfect knowledge of $\rho$.
Further, a short ($l=9$) `non-uniform' sampling based
$\rho$-training sequence designed with ${\mathcal Q}_{(M,l)}^{+} = \{ 1/9, 1/5, 1/4, 4/9, 5/8, 1, 5/3, 8/3, 4\} \subset
{\mathcal Q}_{M}^{+}$ achieves a BER is close to that achieved by the optimal $\rho$-training sequence designed using Theorem \ref{Thm3} (compare
curves 3 and 4).
Also, the BER performance of a `uniform' sampling based $\rho$-training sequence design
with ${\mathcal Q}_{(M,l)}^{+} = \{ 1/9, 8/9, 8/5, 9/4, 3, 4, 5, 8, 9\}$ (the induced intervals are almost
uniformly distributed)  
is inferior to the BER performance achieved by a `non-uniform' sampling based design (compare curves
4 and 5).
Finally, it is observed that, the BER achieved with the $\rho$-training sequence
designed using (\ref{exp_tr_design}) (with $d=1.57, l=9$) is similar to the BER achieved with perfect
knowledge of $\rho$ (compare curves 1 and 6). 
\begin{figure}[t]
\begin{center}
\hspace{-7mm}
\epsfig{file=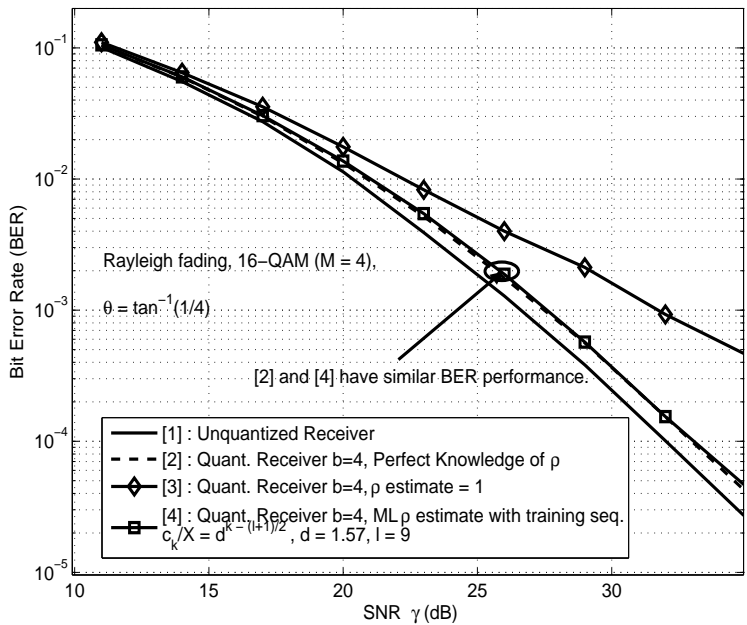, width=100mm,height=85mm}
\end{center}
\vspace{-5mm}
\caption{BER performance with a quantized receiver ($b=4$) and imperfect receiver knowledge of $\rho$. Rotated 16-QAM ($M=4$).}
\vspace{-4mm}
\label{rho_est_perf_b4_16qam}
\end{figure}
For higher order rotated $M^2$-QAM, we proposed `good' $\rho$-training sequences which are short and
have near-optimal performance.
We support this fact through Fig.\ref{rho_est_perf_b4_16qam}, where
we plot the BER performance for a rotated 16-QAM constellation ($\theta = \tan^{-1}(1/4)$), with a quantized receiver ($b =4$) and minimum distance decoding at the receiver with imperfect knowledge of $\rho$. An estimate of $\rho$ is computed based on the
proposed ML estimation scheme discussed in Section \ref{sec_rho_estimate}. The $\rho$-training sequence used for estimation is
the same $\rho$-training sequence used in simulation curve 6 of Fig.\ref{rho_est_perf_b2_4qam}.
From curve 4 in the Fig.\ref{rho_est_perf_b4_16qam}, it is observed that with a short $\rho$-training sequence of only 9 symbols,
it is possible to achieve a BER performance comparable to the BER performance achieved with perfect knowledge
of $\rho$ (curve 2). This is also interesting since the same training sequence was also observed to be near-optimal with $M=2,b=2$. It therefore appears that the length of near-optimal/`good' $\rho$-training sequences does not increase significantly with increasing QAM size. One possible reason for this could be that with increasing QAM size, the quantizer resolution $b$ also increases,
which makes the estimate of $\rho$ more reliable.
\newpage
\section{Conclusions}\label{conclusions}
In this paper, we addressed the problem of achieving modulation diversity in fading channels
with quantized receiver. For 2-dimensional modulation coding, through analysis we showed that in quantized receivers with perfect channel knowledge, algebraic rotations with {\em equidistant} projections can achieve modulation diversity with low complexity minimum distance decoding.
We then relaxed the perfect channel knowledge assumption, and proposed novel channel training/estimation, which were shown to achieve an error probability performance similar to that achieved with perfect channel knowledge.

\newpage
\appendices
\section {A matched rotated constellation achieves full modulation diversity.}\label{sec_Thm_full_div}
The main result that we prove in this section is that, a matched rotated constellation is guaranteed to achieve full modulation diversity.
For the sake of clarity, we formally define error probability and diversity.
The average error probability (averaged over both the fading statistics and the
transmitted vector) is given by
\begin{equation}
\label{pe_eq}
P (\gamma) = \frac {1}{\vert {\mathcal X} \vert}  \sum_{{\bf x} \in {\mathcal X}} P_e({\bf x})
\end{equation}
where it is assumed that all transmit vectors are equiprobable
and $P_e({\bf x})$ denotes the average error probability (averaged over the fading
realizations $ h_1 $ and $ h_2 $) when ${\bf x}$ is the transmitted
vector and minimum distance decoding is performed on the quantized output (\ref{euc_dis_dec}).
Further it is implicitly assumed that $P_e({\bf x})$ is also a function of the SNR
$\gamma$. 
The diversity achieved is given by
\begin{equation}
\label{div_def}
\delta \Define \lim_{\gamma \rightarrow \infty} \frac {\log P(\gamma)} {\log \gamma}.
\end{equation}

Using the union bounding technique, $P(\gamma)$
can be upper bounded as
\begin{equation}
\label{upp_bnd_pe_final}
P(\gamma) \leq \frac {1}{\vert {\mathcal X} \vert} \sum_{{\bf x} \in {\mathcal X}} \sum_{{\bf y} \ne {\bf x}} P_e({\bf x} , {\bf y}).
\end{equation}
where $P_e({\bf x} , {\bf y})$ is the average pairwise error probability
of the event that the minimum distance decoder decodes in favor of ${\bf y}$
when ${\bf x}$ was actually transmitted.

If it can be shown that the pairwise diversity
\begin{equation}
\label{d_x_y}
\delta({\bf x},{\bf y}) \Define \lim_{\gamma \rightarrow \infty} \frac { \log P_e({\bf x}, {\bf y})} {\log \gamma} = 2
\end{equation}
then from the union bound in (\ref{upp_bnd_pe_final}) it follows that the achievable diversity $\delta$ is indeed equal to 2
(since with $n=2$ and i.i.d. Rayleigh fading, the maximum achievable diversity is 2).
The following theorem shows that the statement in (\ref{d_x_y}) is indeed true for a
matched rotated constellation.
\begin{mytheorem}\label{Thm_full_div}
Consider a rotated $M^2$-QAM constellation which is matched
with the $b=2\lceil \log_2(M) \rceil$-bit quantizer.
A minimum distance decoder on the quantized output
achieves $\delta({\bf x},{\bf y}) = 2$ for any two distinct transmit vectors
${\bf x}$ and ${\bf y}$.
\end{mytheorem}
{\em Proof}:
We need to show that the pairwise error probability between
any two transmit vectors has a diversity order of $2$. We only consider the real
components, since the rotation matrix is real valued and therefore the error probability
of the imaginary component would be the same.
This implies that the pairwise diversity in (\ref{d_x_y}) is the same as
that between the real component of the transmitted vectors.  

\begin{figure}[t]
\begin{center}
\hspace{-7mm}
\epsfig{file=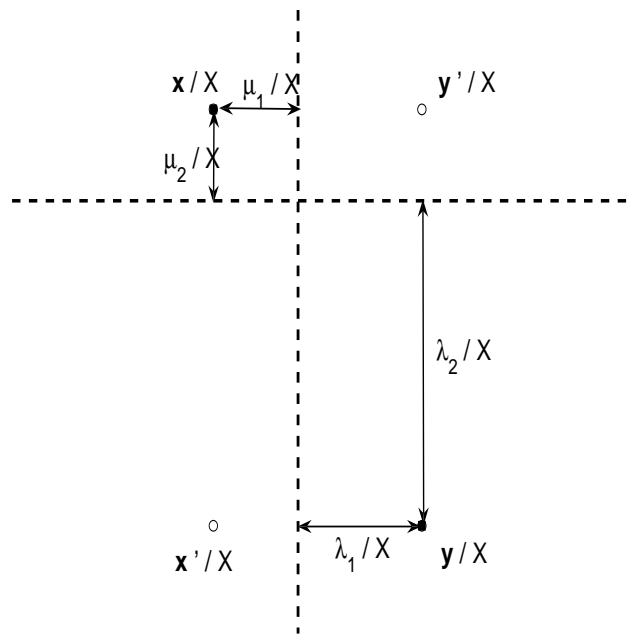, width=85mm,height=85mm}
\end{center}
\vspace{-5mm}
\caption{Reduced system model.}
\vspace{-5mm}
\label{siso_proof_fig}
\end{figure}

Consider any two distinct transmit vectors whose real components are denoted by ${\bf x} = (x_1, x_2)^T$ and ${\bf y}=(y_1,y_2)^T$. Since the projections along both the codeword components are equidistant, ${\bf Q}_b({\bf x}/{X}) =  {\bf x}/{X}$ and the same is true for ${\bf y}$.
Further, it is always possible to find two other vectors
${\bf x}'=(x'_1,x'_2)^T$, and ${\bf y}'= (y'_1,y'_2)^T$ such that, $x'_1 = x_1$, $x'_2 = y_2$, $y'_1 = y_1$ and $y'_2 = x_2$.
It is also noted that,
${\bf Q}_b({\bf x}'/{X}) = {\bf x}'/{X}$, and the same is true for ${\bf y}'$.
Effectively, ${\bf x}/X$, ${\bf y}/X$, ${\bf x}'/X$ and ${\bf y}'/X$ form the four vertices of a rectangle (Fig. \ref{siso_proof_fig}). We next focus only on this rectangle, and consider the constellation with only 2 possible transmit vectors, namely ${\bf x}$ and ${\bf y}$.
Therefore, we now
have a reduced system in which only 1 bit is communicated (since we have only 2 possible transmit vectors). Also, for the reduced system we design a {\em non-uniform} quantizer such that the four vertices of the rectangle correspond
to the four possible quantized outputs (Fig. \ref{siso_proof_fig}).
As shown in Fig. \ref{siso_proof_fig}, the received signal space is partitioned into
four quantization regions by a quantization boundary in each component (depicted in the figure
by dashed lines, there is one quantization boundary along the horizontal component and another along the
vertical component). 
We are specially interested in investigating the {\em worst possible} pairwise error probability
when ${\bf x}$ is transmitted and the minimum distance decoder decodes in favor of ${\bf y}$.
Since we are interested in the worst case, we choose the two quantization boundaries (one for each component) of the reduced system to be the same as those two quantization boundaries in the {\em original system} which separate ${\bf x}$ and ${\bf y}$ along both the components in the original system and are also closest to ${\bf x}$ in the {\em original system}.

If the error probability of the reduced system has a diversity order of 2, then since the original system has more quantized output points than the reduced system (in addition to the four points of the reduced system), it is obvious
that the error probability between ${\bf x}$ and ${\bf y}$ in the original system would always be smaller than that in the reduced system.
Therefore with this assumption, it suffices to show that the diversity order of $P_e({\bf x},{\bf y})$ in the reduced system is 2.
For the reduced system there are two possible cases, either {\em i}) ${\bf x}/X$ and ${\bf y}/X$ are at the positions shown in Fig.~\ref{siso_proof_fig}, or {\em ii}) the positions of ${\bf x}/X$ and ${\bf y}/X$ are swapped with each other. Further for each case, the positions of ${\bf x}/X$ and ${\bf y}/X$ could also be swapped with the positions of ${\bf x}'/X$ and ${\bf y}'/X$ respectively. This therefore results in four possible scenarios, out of which we only discuss the scenario shown in Fig.\ref{siso_proof_fig}, since the analysis and pairwise diversity
for the other three scenarios is same.
Out of the four possible scenarios, let us denote the scenario shown in Fig. \ref{siso_proof_fig} by the anti-clockwise ordered
tuple starting from the first quadrant i.e., $({\bf y}',{\bf x},{\bf x}',{\bf y})$.
With this notation, the other three possible scenarios
are given by the tuples $({\bf x}',{\bf y},{\bf y}',{\bf x})$, $({\bf y},{\bf x}',{\bf x},{\bf y}')$ and
$({\bf x},{\bf y}',{\bf y},{\bf x}')$.

Lets consider that ${\bf x}$ is the transmitted vector in the reduced system. For the minimum distance decoder
to decode in favor of ${\bf y}$, the received symbol vector $(r_1^I,r_2^I)$ must lie in a quantization
box whose quantized output is closer to ${\bf y}$ in terms of the Euclidean distance $d_E(.,.)$ defined below. The Euclidean distance between any two vectors ${\bf u}$ and
${\bf v}$ is defined as
\begin{equation}
\label{de_eqn}
d_E({\bf u},{\bf v}) \Define {\Bigg (} {\Big (} \frac{{\bf G}{\bf u}}{X} {\Big )} - {\Big (} \frac{{\bf G}{\bf v}}{X} {\Big )} {\Bigg )}^{T} {\bf D}_{\rho} {\Bigg (} {\Big (} \frac{{\bf G}{\bf u}}{X} {\Big )} - {\Big (} \frac{{\bf G}{\bf v}}{X} {\Big )} {\Bigg )}.
\end{equation}
From Fig. \ref{siso_proof_fig} it is clear that when $\vert h_2 \vert (\mu_2 + \lambda_2) > \vert h_1 \vert (\mu_1 + \lambda_1)$, $d_E({\bf y}',{\bf x}) < d_E({\bf y}',{\bf y})$ and $d_E({\bf x}',{\bf y}) < d_E({\bf x}',{\bf x})$. That is, the quantized output ${\bf y}'/X$ is closer to ${\bf x}/X$ than to ${\bf y}/X$, and ${\bf x}'/X$ is closer to ${\bf y}/X$ in terms of Euclidean distance.
Therefore, the minimum distance decoder decodes in favor of ${\bf y}$ only when the received symbol vector
lies in the quantization box corresponding to either ${\bf x}'/X$ or ${\bf y}/X$.
This in turn happens only when the noise along the second component is less than $-\vert h_2 \vert \mu_2$. 
Therefore, when $\vert h_2 \vert (\mu_2 + \lambda_2) > \vert h_1 \vert (\mu_1 + \lambda_1)$, a decoding error occurs if and only if, the noise along the second component ($w_2^I$) is less than $-\vert h_2 \vert \mu_2$.

Similarly, when
$\vert h_2 \vert (\mu_2 + \lambda_2) < \vert h_1 \vert (\mu_1 + \lambda_1)$, a decoding error occurs if and only if, the noise along the first component ($w_1^I$) is more than $\vert h_1 \vert \mu_1$.
These two error events are summarized as
\begin{eqnarray}
\label{error_events}
E_1 &:& w_2^I < -\vert h_2 \vert \mu_2\,,\,  \mbox{and} \,\,\vert h_2 \vert (\mu_2 + \lambda_2) > \vert h_1 \vert (\mu_1 + \lambda_1) \nonumber \\
E_2 &:& w_1^I > \vert h_1 \vert \mu_1\,,\, \mbox{and} \,\, \vert h_2 \vert (\mu_2 + \lambda_2) < \vert h_1 \vert (\mu_1 + \lambda_1) \nonumber \\
\end{eqnarray}
The error events can be written equivalently as
\begin{eqnarray}
\label{equiv_error_events}
E_1 &:& {\Bigg \{} w_2^I < -\mu_2 \max {\Bigg (} \vert h_2 \vert \,,\, \vert h_1 \vert \frac {(\mu_1 + \lambda_1)}{(\mu_2 + \lambda_2) } {\Bigg )} \nonumber \\
      & &  \,\,\, \mbox{and} \,\,\vert h_2 \vert (\mu_2 + \lambda_2) > \vert h_1 \vert (\mu_1 + \lambda_1) {\Bigg \}}\nonumber \\
E_2 &:& {\Bigg \{} w_1^I > \mu_1 \frac {(\mu_2 + \lambda_2)}{(\mu_1 + \lambda_1)} \max {\Bigg (} \vert h_2 \vert \,,\, \vert h_1 \vert \frac {(\mu_1 + \lambda_1)}{(\mu_2 + \lambda_2) } {\Bigg )} \nonumber \\
& & \,\,\,  \mbox{and} \,\,\vert h_2 \vert (\mu_2 + \lambda_2) < \vert h_1 \vert (\mu_1 + \lambda_1) {\Bigg \}}.
\end{eqnarray}

Let $\mbox{Pr}(E)$ be used to denote the probability of some event $E$.
$P_e({\bf x},{\bf y})$ can therefore be  expressed as\footnote{\footnotesize {With a slight abuse of notation, we still use $P_e({\bf x},{\bf y})$ to denote the pairwise error probability in the {\em reduced system}.}}
\begin{eqnarray}
\label{pe_expr}
P_e({\bf x}, {\bf y}) &=& {\mathbb  E}_{ h_1 ,h_2 } {\Big [} \mbox{Pr}(E_1 \cup E_2) {\Big ]} \nonumber \\
&=& {\mathbb  E}_{ h_1 , h_2 } {\Big [} \mbox{Pr}(E_1) +  \mbox{Pr}(E_2) {\Big ]}
\end{eqnarray}
where the second statement follows from the fact that $E_1$ and $E_2$ have no common support (i.e., $E_1 \cap E_2 = \phi$).
We further define two more events
\begin{eqnarray}
\label{error_events_tilde}\footnotesize
\Tilde{E_1} &:& {\Bigg \{} -w_2^I > \min {\Big (} \mu_2, \mu_1 \frac {(\mu_2 + \lambda_2)}{(\mu_1 + \lambda_1)} {\Big )} \max {\Bigg (} \vert h_2 \vert \,,\, \vert h_1 \vert \frac {(\mu_1 + \lambda_1)}{(\mu_2 + \lambda_2) } {\Bigg )} \nonumber \\
& & \,\,\,  \mbox{and} \,\,\vert h_2 \vert (\mu_2 + \lambda_2) > \vert h_1 \vert (\mu_1 + \lambda_1)  {\Bigg \}} \nonumber \\
\Tilde{E_2} &:& {\Bigg \{} w_1^I >  \min {\Big (} \mu_2, \mu_1 \frac {(\mu_2 + \lambda_2)}{(\mu_1 + \lambda_1)} {\Big )} \max {\Bigg (} \vert h_2 \vert \,,\, \vert h_1 \vert \frac {(\mu_1 + \lambda_1)}{(\mu_2 + \lambda_2) } {\Bigg )} \nonumber \\
& &  \,\,\,  \mbox{and} \,\,\vert h_2 \vert (\mu_2 + \lambda_2) < \vert h_1 \vert (\mu_1 + \lambda_1) {\Bigg \}}.
\end{eqnarray}
Since ${E_1} \subset \Tilde{E_1}$ and $E_2 \subset \Tilde{E_2}$ it follows that
\begin{equation}
\label{pe_bnd}
P_e({\bf x},{\bf y}) \leq {\mathbb  E}_{h_1 , h_2 } {\Big [} \mbox{Pr}(\Tilde{E_1}) +  \mbox{Pr}(\Tilde{E_2}) {\Big ]}.
\end{equation}

A careful inspection of the error events $\Tilde {E_1}$ and $\Tilde {E_2}$ reveals that
in both cases, we are interested in an event when a zero mean and finite variance Gaussian random variable
is greater than the same constant (i.e., $\min {\Big (} \mu_2, \mu_1 \frac {(\mu_2 + \lambda_2)}{(\mu_1 + \lambda_1)} {\Big )}$).
Therefore, using (\ref{error_events_tilde}) and (\ref{pe_bnd}), we have
\begin{eqnarray}
\label{pe_defs}
P_e({\bf x},{\bf y}) & \leq & {\mathbb  E}_{ h_1 , h_2} {\Big [} \mbox{Pr} (w > \beta \sqrt{z})  {\Big ]}  \nonumber \\
w  & \sim & {\mathcal N}{\Bigg (} 0, \frac{\sigma^2}{2} {\Bigg )} \nonumber \\
\beta & \Define &  \min {\Bigg (} \mu_2, \mu_1 \frac {(\mu_2 + \lambda_2)}{(\mu_1 + \lambda_1)} {\Bigg )} \nonumber \\
z     & \Define &  \max {\Bigg (} \vert h_2 \vert^2 \,,\, \alpha \vert h_1 \vert^2 {\Bigg )} \nonumber \\
\alpha & \Define & \frac {(\mu_1 + \lambda_1)^2}{(\mu_2 + \lambda_2)^2 }.
\end{eqnarray}
Then
\begin{equation}
\label{gaussian_error_pe}
\mbox{Pr} (w > \beta \sqrt{z}) = \Phi {\Bigg (} \sqrt{\frac{2 \beta^2 {z}}{\sigma^2}} {\Bigg )}
\end{equation}
where $\Phi(x) \Define \frac{1}{\sqrt{2 \pi}} \int_{x}^{\infty} e^{-\frac{t^2}{2}} dt$.
Using (\ref{gaussian_error_pe}) and (\ref{pe_defs}), we have
\begin{equation}
\label{new_pe_bnd}
P_e({\bf x},{\bf y})  \leq  {\mathbb  E}_{z} {\Bigg [} \Phi {\Bigg (} \sqrt{\frac{2 \beta^2 {z}}{\sigma^2}} {\Bigg )}  {\Bigg ]}.
\end{equation}

With a simple algebraic manipulation on the right hand side term in (\ref{new_pe_bnd}), it can be shown that
\begin{equation}
\label{reduced_pe_1_2_sisoF}
P_e({\bf x},{\bf y}) \leq \frac{1}{\sqrt{2 \pi}} \int_{0}^{\infty} F_{z} {\Big (}\frac{x^2\sigma^2}{2 \beta^2} {\Big )} e^{-\frac{x^2}{2}} dx,
\end{equation}
where $F_{z}(x) \Define \mbox{Pr}({z} \leq x)$ is the cumulative density function of ${z}$.
Under the Rayleigh fading assumption, $\vert h_1 \vert^2$ and $\vert h_2 \vert^2$ are i.i.d. exponential random variables with mean 1, and hence
$F_{z}(x)= (1 - e^{-z}) (1 - e^{-\frac{z}{\alpha}})$.
Integrating the right hand side term in (\ref{reduced_pe_1_2_sisoF}), we obtain
\begin{equation}
\label{reduced_pe_12_final}
P_e({\bf x},{\bf y}) \leq  \frac{1}{2} {\Bigg \{} 1 - \frac{1}{\sqrt{1 + \frac{\sigma^2}{\beta^2} }} - \frac{1}{\sqrt{1 + \frac{\sigma^2}{\alpha \beta^2} }} +  \frac{1}{\sqrt{1 + \frac{\sigma^2}{\beta^2}  (1 + \frac {1}{\alpha})}} {\Bigg \} }.
\end{equation}
As $\gamma \rightarrow \infty$ the series expansion of
(\ref{reduced_pe_12_final}) yields
\begin{eqnarray}
\label{pe_taylor}
P_e({\bf x},{\bf y}) & \leq & \hspace{-3mm} \frac{3 P_T^2}{8\alpha \beta^4} \gamma^{-2} + o(\gamma^{-3}) \nonumber \\
& \leq & \hspace{-3mm} \frac{3 P_T^2}{8 \alpha \min {\Big (} \mu_2^4 , \mu_1^4 \frac {(\mu_2 + \lambda_2)^4}{(\mu_1 + \lambda_1)^4}  {\Big )} } \gamma^{-2} + o(\gamma^{-3}) 
\end{eqnarray}
where $o(.)$ denotes the little-o notation\footnote{Any function $f(x)$ in a single variable $x$ is said to be $o(g(x))$  i.e., $f(x) = o(g(x))$ if ${\frac {f(x)} {g(x)}} \rightarrow 0$ as ${x \rightarrow 0}$.}.

Since the projections along both the codeword components are distinguishable and equidistant,
$\mu_1$, $\mu_2$, $\lambda_1$ and $\lambda_2$ (and therefore $\alpha$ and $\beta$) are non-zero positive quantities for any pair of transmit vectors $({\bf x}$,${\bf y})$.
The asymptotic expressions in (\ref{pe_taylor}) reveal that the pairwise error probability $P_e({\bf x},{\bf y})$ indeed achieves second order diversity (i.e., $\delta({\bf x},{\bf y}) = 2$).
$\hfill \blacksquare$
\begin{figure}[t]
\begin{center}
\hspace{-7mm}
\epsfig{file=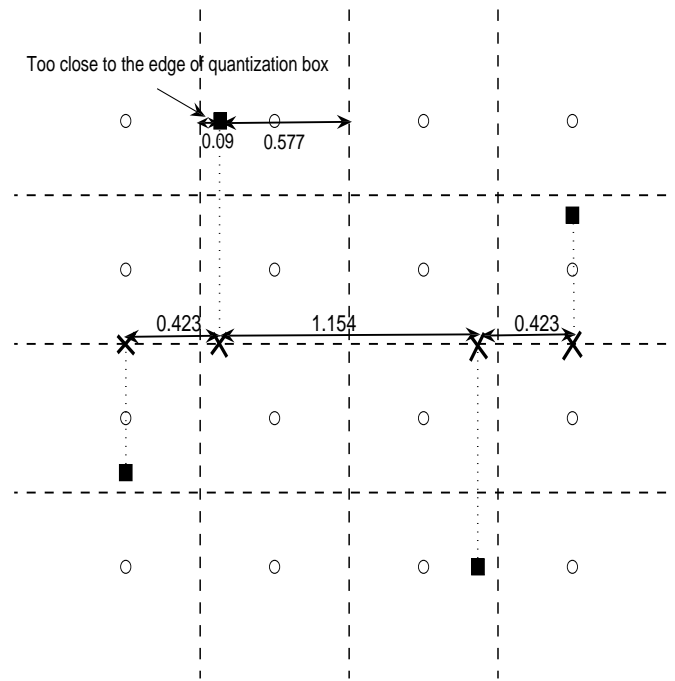, width=90mm,height=90mm}
\end{center}
\vspace{-5mm}
\caption{Example of a rotated constellation that is mismatched with the quantizer (i.e., non-equidistant projections).
$b=2$, $\theta=\pi/12$ and $4$-QAM.}
\label{fig_non_equidistant}
\end{figure}

We next discuss the reason why a `good' choice of the rotation angle must be
one which results in a matched rotated constellation.
Through Fig.\ref{fig_non_equidistant}, we illustrate the reasoning behind
the fact that a rotated constellation matched with the quantizer (i.e., with equidistant projections) achieves
a lower error probability when compared to a mismatched rotated constellation. 
We plot the
four transmit vectors and the 16-possible quantized outputs, for a mismatched rotated 4-QAM constellation ($\theta = \pi/12$), and a $b=2$-bit quantizer.
The projections of the transmit vectors onto the
horizontal component (marked with a cross) are {\em not} equidistant (distances between the
projections are $0.423$, $1.154$ and $0.423$). Due to this, the transmit vectors are not at the centre
of their respective quantization boxes. This then implies that a transmit vector is closer to
one edge of its quantization box compared to the other edges. Therefore for a given noise variance,
when compared to a rotation code with equidistant projections (i.e., a matched rotated constellation),
a transmit vector in a code with non-equidistant projections has a {\em higher probability} of being received in a
different quantization box whose quantized output is closer to some other transmit vector.
\begin{figure}[t]
\begin{center}
\hspace{-7mm}
\epsfig{file=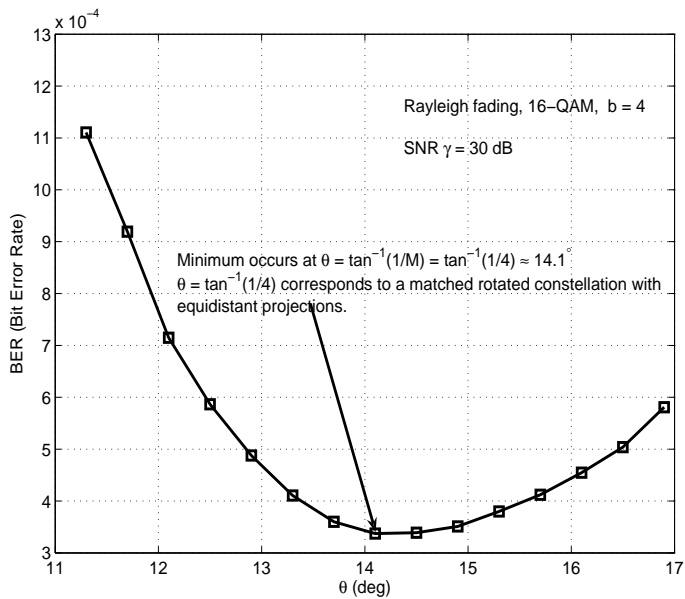, width=95mm,height=80mm}
\end{center}
\vspace{-5mm}
\caption{BER of rotated 16-QAM at SNR $\gamma=30$ dB and varying $\theta$ in the range of admissible angles $(11.3^{\circ}, 16.9^{\circ}  )$. $b=4$ and minimum distance decoding.}
\vspace{-4mm}
\label{siso_qam_crit3}
\end{figure}

Through the simulation plots in Fig.\ref{siso_qam_crit3}, we experimentally support the validity of the fact that a rotated-QAM
constellation matched with the quantizer has a lower error probability than a mismatched rotated
constellation.
The BER performance of a rotated 16-QAM modulation code is plotted for varying rotation angle $\theta$.
The SNR $\gamma$ is fixed to 30 dB and the rotation angle is constrained to lie within
the range of admissible angles, i.e., $(11.3^{\circ},16.9^{\circ})$ for 16-QAM.
Minimum distance decoding is performed on the quantized output ($b=4$).
It is observed that the best BER performance is attained at $\theta = \tan^{-1}(1/4)$
which corresponds to a matched rotated constellation having component-wise equidistant projections. Hence, this observation clearly validates
code design Criterion III.
\section {Theorem \ref{Thm_equi_code}}\label{sec_Thm_equi_code}
\begin{mytheorem}\label{Thm_equi_code}
With $M^2$-QAM information symbols, the rotation code with $\theta = \tan^{-1}(1/M)$ satisfies Criterion III. 
\end{mytheorem}

{\em Proof}:
The corresponding rotation matrix ${\bf G}$, with
$\theta = \tan^{-1}(1/M)$ is given by
\begin{equation}
\label{construct_K2}
{\bf G} = \frac{1}{\sqrt{M^2 + 1}}\left[\begin{array}{cc}
M & 1 \\
-1 &  M
\end{array} \right]
\end{equation}
Since ${\bf G}$ is real-valued, it suffices to prove the equidistant projections
property for the real component only.
The information symbols $u_1^I$ and $u_2^I$ take values from the $M$-PAM
signal set ${\mathcal S}_M$.
It is easy to show that with ${\bf G}$ given by (\ref{construct_K2}), there are $M^2$ distinct equidistant projections on both the
codeword components. The pair
$(u_1^I, u_2^I)$ can take any of the $M^2$ values from the ordered sequence of
values $ {\mathcal S}_M^{1} = \{  (-M+1,-M+1), (-M+1,-M+3), \ldots (-M+1,M-1), \, (-M+3,-M+1), (-M+3,-M+3), \ldots (-M+3,M-1), \ldots (M-1,-M+1), (M-1,-M+3), \ldots (M-1,M-1) \}$.
It is easy to see that the value of the first component of the transmit vector, $x_1^I$, increases in steps of $2/\sqrt{M^2 + 1}$ as $(u_1^I,u_2^I)$ takes values sequentially from the set
${\mathcal S}_M^{1}$. This then proves that the projections along the first component are indeed
equidistant.

Also, since the values in the $M$-PAM signal set
are symmetric around 0, it follows that the set of all $M^2$
values which the first component $x_1^I$ takes, is same as the set of the $M^2$
values taken by the second component $x_2^I$.
Hence the projections along the second component are also equidistant.
$\hfill \blacksquare$
\section {Proof of Theorem \ref{Thm1}}\label{sec_Thm1}
For a given channel realization, let the received vector on the real component of the
channel be ${\bf r}^I$. From (\ref{euc_dis_dec_IQ_1}), it is obvious that
\begin{equation}
\label{thm1_proof_eq1}
m(\rho,{\bf r}^I,{\widehat {\bf u}}^I) \leq  m(\rho,{\bf r}^I,{\Tilde {\bf u}}^I)
\end{equation}
for any information symbol vector ${\Tilde {\bf u}}$.
Using the definition of $D_{E}(\cdot,\cdot,\cdot,\cdot)$ in (\ref{D_dist_def}),
(\ref{thm1_proof_eq1}) can be written as
\begin{equation}
\label{thm1_proof_eq2}
D_{E}(\rho,{\bf r}^I,{\widehat {\bf u}}^I,{\Tilde {\bf u}}^I) \leq  0
\end{equation}
for any information symbol vector ${\Tilde {\bf u}}$.

Let us assume that, the estimate of $\rho$, ${\hat \rho}$, satisfies (\ref{thm1_statement})
for all received vector ${\bf r}$, and all information
symbol vectors ${\bf u}$ and ${\bf v}$. Hence, ${\hat \rho}$ must
satisfy (\ref{thm1_statement}) for ${\bf u} = {\widehat {\bf u}}$
and ${\bf v} = {\Tilde {\bf u}}$. This, then implies that
\begin{equation}
\label{thm1_proof_eq3}
D_{E}(\rho,{\bf r}^I,{\widehat {\bf u}}^I,{\Tilde {\bf u}}^I) D_{E}({\hat \rho},{\bf r}^I,{\widehat {\bf u}}^I,{\Tilde {\bf u}}^I) \geq 0
\end{equation}
for any information symbol vector ${\Tilde {\bf u}}$.

Combining (\ref{thm1_proof_eq2}) and (\ref{thm1_proof_eq3}) we have
\begin{equation}
\label{thm1_proof_eq4}
D_{E}({\hat \rho},{\bf r}^I,{\widehat {\bf u}}^I,{\Tilde {\bf u}}^I) \leq  0
\end{equation}
for any information symbol vector ${\Tilde {\bf u}}$.
This however implies that
\begin{equation}
\label{thm1_proof_eq5}
m({\hat \rho},{\bf r}^I,{\widehat {\bf u}}^I) \leq  m({\hat \rho},{\bf r}^I,{\Tilde {\bf u}}^I)
\end{equation}
for any information symbol vector ${\Tilde {\bf u}}$.
This then means that the output of the minimum distance decoder with the imperfect estimate ${\hat \rho}$,
is indeed the same as the output of the minimum distance decoder assuming perfect receiver knowledge of $\rho$.
Since, this is true for any received vector ${\bf r}$, it is obvious that the error probability
performance of the minimum distance decoder with imperfect $\rho$ estimate is the same as the error
probability performance with perfect knowledge of $\rho$.
Hence, any estimate of $\rho$, which satisfies (\ref{thm1_statement}) for all ${\bf r}$,${\bf u}$ and ${\bf v}$, is indeed an {\em optimal} estimate.
$\hfill \blacksquare$
\section {Proof of Theorem \ref{Thm2}}\label{sec_Thm2}
We basically show that if (\ref{thm2_statement}) is satisfied for all $l \in {\mathcal Q}_M^{+}$, then this implies that (\ref{thm1_statement}) is satisfied for all possible received vector ${\bf r}$ and
information symbol vectors ${\bf u}$ and ${\bf v}$. The optimality of ${\hat \rho}$ then follows
from the application of Theorem \ref{Thm1}.

For any given received vector ${\bf r}$, and information symbol vectors ${\bf u}$ and ${\bf v}$, assuming
(\ref{thm2_statement}), we would eventually show that the given ${\bf r}$, ${\bf u}$ and ${\bf v}$
satisfy the sufficiency condition in (\ref{thm1_statement}).
For a given ${\bf r}$, ${\bf u}$ and ${\bf v}$, let $d_1$ and $d_2$ be defined as
\begin{eqnarray}
\label{lm_def}
d_1 \Define (r_1^I - \frac{x_1^I}{X})^2 - (r_1^I - \frac{y_1^I}{X})^2 \nonumber \\
d_2 \Define (r_2^I - \frac{x_2^I}{X})^2 - (r_2^I - \frac{y_2^I}{X})^2
\end{eqnarray}
where
\begin{eqnarray}
\label{tx_vecs}
{\bf x} = ( x_1 , x_2)^T = {\bf G} {\bf u} \,,\, {\bf y} = (y_1 , y_2)^T = {\bf G} {\bf v}.
\end{eqnarray}

It is noted here that, since the transmitted vectors have the equidistant projections property,
$(r_1^I - \frac{x_1^I}{X}), (r_1^I - \frac{y_1^I}{X}), (r_2^I - \frac{x_2^I}{X}), (r_2^I - \frac{y_2^I}{X}) \in {\mathcal D}_M$.
This then implies that, when $d_2 \ne 0$,
$d_1 / d_2 \in {\mathcal Q}_M$.
Further, $D_{E}(\zeta,{\bf r}^I,{\bf u}^I,{\bf v}^I)$ can now be written in terms of
$d_1$ and $d_2$ as follows
\begin{equation}
\label{D_dist_d1d2}
D_{E}(\zeta,{\bf r}^I,{\bf u}^I,{\bf v}^I) = d_1 + {\zeta}^2 d_2.
\end{equation}

We now distinguish between two important events, depending on
whether $d_2=0$ or $d_1=0$ or both are non-zero.
We first consider the case when at least one among $d_1$ and $d_2$ is zero.
If $d_2=0$, then using (\ref{D_dist_d1d2}) it is trivially true that
$D_{E}({\rho},{\bf r}^I,{\bf u}^I,{\bf v}^I)D_{E}({\hat \rho},{\bf r}^I,{\bf u}^I,{\bf v}^I) = d_1^2 \geq 0$, thereby satisfying the sufficiency condition in (\ref{thm1_statement}).
Similarly, when $d_1 = 0$, it is again trivially true that
$D_{E}({\rho},{\bf r}^I,{\bf u}^I,{\bf v}^I)D_{E}({\hat \rho},{\bf r}^I,{\bf u}^I,{\bf v}^I) = d_2^2 \rho^2 {\hat \rho}^2 \geq 0$, thereby satisfying the sufficiency condition in (\ref{thm1_statement}).

When $d_2\ne 0$ and $d_1 \ne 0$, we consider two cases depending on the sign of $d_1$ and $d_2$.
If both $d_1$ and $d_2$ have the same sign, then $sign(d_1 + \rho^2 d_2) = sign(d_1) = sign(d_1 + {\hat \rho}^2 d_2)$, and therefore the sufficiency condition in (\ref{thm1_statement}) is satisfied.
Hence, the only case which remains to be handled is when $sign(d_1) = -sign(d_2)$.

When $sign(d_1) = -sign(d_2)$, we have two possible situations, depending upon the sign of $(d_1 + \rho^2 d_2)$.
When $(d_1 + \rho^2 d_2) > 0$ and $d_1 > 0, d_2 < 0$, it follows that
$\rho^2 < (\vert d_1 \vert / \vert d_2 \vert)$.  Since $(\vert d_1 \vert / \vert d_2 \vert) > 0$ and
$(\vert d_1 \vert / \vert d_2 \vert) \in {\mathcal Q}_M^{+}$, from (\ref{thm2_statement}) it follows that
${\hat \rho}^2  < (\vert d_1 \vert / \vert d_2 \vert)$. Since, $d_1 > 0, d_2 < 0$, this then implies
that $(d_1 + {\hat \rho^2} d_2) > 0$, and therefore $D_{E}({\rho},{\bf r}^I,{\bf u}^I,{\bf v}^I)D_{E}({\hat \rho},{\bf r}^I,{\bf u}^I,{\bf v}^I) = (d_1 + \rho^2 d_2)(d_1 + {\hat \rho^2} d_2) > 0$, thereby
satisfying the sufficiency condition in (\ref{thm1_statement}).

If $(d_1 + \rho^2 d_2) > 0$ and $d_1 < 0, d_2 > 0$, it follows that $\rho^2 > (\vert d_1 \vert / \vert d_2 \vert)$.
From (\ref{thm2_statement}), it then follows that
${\hat \rho}^2  > (\vert d_1 \vert / \vert d_2 \vert)$. Since $d_1 < 0, d_2 > 0$, $(d_1 + {\hat \rho^2} d_2) > 0$, thereby satisfying the sufficiency condition in (\ref{thm1_statement}).

In a similar way, if $(d_1 + \rho^2 d_2) < 0$, then also it can be shown that the sufficiency condition in (\ref{thm1_statement}) is satisfied.
$\hfill \blacksquare$
\section {Proof of Theorem \ref{Thm3}}\label{sec_Thm3}
For the $\rho$-training sequence given by (\ref{thm3_statement}) and the proposed ML estimation scheme,
we would like to show that the ML interval of each {\em feasible} output sequence is exactly same as some interval induced by the set ${\mathcal Q}_M^{+}$.

From (\ref{thm3_statement}) it is obvious that, since $\{ q_k \}$ is an increasing sequence w.r.t. increasing $k$,
$\{ c_k \}$ is also an increasing sequence.
Since the uniform quantizer function $Q_b(.)$ given by (\ref{qri}) is also a monotonically non-decreasing function
of its argument, it follows that {\em i}) the output sequence $\{ r_k \}$ is also a non-decreasing sequence
, and {\em ii}) since $\{ c_k \}$ is a positive valued sequence, it follows that $\{ r_k \}$ is also
a positive valued sequence. For a $b=2$-bit quantizer, the only possible positive values that can be taken by
the quantizer output are $\{ 1/3, 1\}$. These properties therefore imply that, the set of all feasible output sequences
must be a subset of the set consisting of the following $L_M+1$ distinct sequences.
\begin{eqnarray}
\label{feasible_seq}
r_1 = r_2 = \cdots = r_{L_M} = 1, \nonumber \\
r_1 = \cdots = r_k = \frac {1}{3}, r_{k+1} = \cdots = r_{L_M} = 1, \,\,\, k=1,2, \cdots {L_M - 1} \nonumber \\
r_1 = r_2 = \cdots = r_{L_M} = \frac {1}{3}.
\end{eqnarray}

We will however show that, in fact, the set of all feasible output sequences is actually same as
the set of sequences given by (\ref{feasible_seq}).
For each sequence in (\ref{feasible_seq}), we derive the ML interval corresponding to it being a possible output sequence.
We see that each ML interval is non-empty and coincides with exactly one of the intervals
induced by ${\mathcal Q}_M^{+}$. This would then imply that, firstly, each output sequence in ({\ref{feasible_seq})
is a feasible output sequence (since its ML interval is non-empty), and secondly, the estimate of $\rho$
is {\em optimal} (since the ML interval of each feasible output sequence coincides with one of the intervals induced by ${\mathcal Q}_M^{+}$, and therefore from (\ref{rho_hat_optimality}) it follows that ${\hat \rho}$ is optimal).

We now derive the ML interval corresponding to each output sequence in (\ref{feasible_seq}).
Using (\ref{segment_k}), the ML interval corresponding to the output sequence $\{ r_1 = r_2 = \cdots = r_{L_M} = 1 \}$ is $[2X / 3c_1  , \infty)$. From (\ref{thm3_statement}), $c_1 = 2X / 3q_{L_M} $ and therefore, the ML interval for the all ones
output sequence is $[ q_{L_M} , \infty)$, which coincides with the last interval of ${\mathcal Q}_M^{+}$.
Similarly, the ML interval corresponding to the output sequence $r_1 = r_2 = \cdots = r_{L_M} = \frac {1}{3}$
is $[ 0 , 2X / 3c_{L_M})$. From (\ref{thm3_statement}), $c_{L_M} = 2X / 3q_{1} $ and therefore,
the ML interval corresponding to the output sequence $r_1 = r_2 = \cdots = r_{L_M} = \frac {1}{3}$ is
$[0 , q_1)$, which coincides with the first interval induced by ${\mathcal Q}_M^{+}$.

For the $k$-th output sequence $r_1 = \cdots = r_k = \frac {1}{3}, r_{k+1} = \cdots = r_{L_M} = 1$,
using (\ref{segment_k}), the ML interval is given by $[2X/3c_{k+1} , 2X/3c_k) = [q_{L_M - k} , q_{L_M - k + 1} )$, which corresponds to one of the intervals induced by ${\mathcal Q}_M^{+}$.
This then proves the optimality of the $\rho$ estimate derived using the proposed ML estimation
scheme with the $\rho$-training sequence given by (\ref{thm3_statement}).
$\hfill \blacksquare$

\begin{thebibliography}{99}
\bibitem{Walden} R. H.\ Walden,
``Analog-to-Digital Converter Survey and Analysis,''
{\em IEEE Journal on Selected Areas in Communications}, pp.\ 539--550, vol.\ 17, no.\ 4, April \ 1999.

\bibitem{Nossek} A.\ Mezghani, M. S.\ Khoufi, and J. A.\ Nossek,
``Maximum Likelihood Detection for Quantized MIMO Systems,''
{\em The International ITG Workshop on Smart Antennas, (WSA'2008)}, pp.\ 278--284, Darmstadt, Germany, Feb \ 2008.

\bibitem{Gareth} M.\ Gareth and A.\ Sabharwal,
``On the Impact of Finite Receiver Resolution in Fading Channels,''
{\em The Forty-Fourth Allerton Conference on Communication, Control and Computing (Allerton'2006)}, Allerton House, UIUC, Illinois, USA, Sept. \ 2006.

\bibitem{Ivrlac} M. T.\ Ivrlac and J. A.\ Nossek,
``Capacity and Coding for Quantized MIMO Systems,''
{\em IEEE International Wireless Communications and Mobile Computing Conference, (IWCMC'2006)}, pp.\ 1387--1392, Vancouver, Canada, July \ 2006.

\bibitem{Singh} J.\ Singh, O.\ Dabeer and U.\ Madhow,
``Capacity of the Discrete-time AWGN Channel Under Output Quantization,''
{\em IEEE International Symposium on Information Theory, (ISIT'2008)}, pp.\ 1218--1222, Toronto, Canada, July \ 2008.

\bibitem{Vit98} J.\ Boutros and E.\ Viterbo,
``Signal Space Diversity: A Power and Bandwidth Efficient Diversity Technique for the Rayleigh Fading Channel,''
{\em IEEE Trans. on Information Theory}, pp.\ 1453--1467, vol.\ 44, no.\ 4, July \ 1998.

\bibitem{Vit96} J. Boutros, E. Viterbo, C. Rastello, and J. C.\ Belfiore, ``Good Lattice Constellations for both Rayleigh Fading and Gaussian Channels," {\em IEEE Trans. on Information Theory}, vol. 42, no. 2, pp. 502--518, March 1996.

\bibitem{QizhengGu} Qizheng Gu, {\em RF System Design of Transceivers for Wireless Communications}, Springer, 1st Ed., 2005.

\bibitem{Liu} S.\ I.\ Liu and C.\ C.\ Chang,
``CMOS Analog Divider and Four-Quadrant Multiplier Using Pool Circuits,''
{\em IEEE Journal of Solid-State Circuits}, pp.\ 1025--1029, vol.\ 30, no.\ 9, Sept.\ 1995.

\bibitem{Vit04} E. Bayer-Fluckiger, F. Oggier, and E. Viterbo, ``New Algebraic Constructions of Rotated ${\mathbb Z}^n$-Lattice Constellations for the Rayleigh Fading Channel," {\em IEEE Trans. on Information Theory}, vol. 50, no. 4, pp. 702--714, Apr. 2004.
\end{thebibliography}
\end{document}